\documentclass[]{emulateapj}

\usepackage{CJK}
\usepackage{amsmath}
\usepackage{bm}
\usepackage{natbib}
\usepackage{mathrsfs}
\usepackage{epstopdf}
\usepackage{txfonts}
\usepackage{graphicx,color}

\newcommand{\rH}{r_{\rm H}}
\newcommand{\rB}{r_{\rm B}}
\newcommand{\rs}{r_{\rm s}}
\newcommand{\cs}{c_{\rm s}}

\newcommand{\dr}{{\rm d}r}
\newcommand{\eqnref}[1]{Equation \ref{#1}}
\newcommand{\secref}[1]{Section \ref{#1}}
\newcommand{\figref}[1]{Figure \ref{#1}}

\newcommand{\ddr}[1]{\frac{{\rm d}#1}{\dr}}
\newcommand{\ddlnr}[1]{\frac{{\rm d}#1}{{\rm d}\ln{r}}}

\shorttitle{3D Simulations of Low Mass Planets}
\shortauthors{Fung, Artymowicz \& Wu}

\begin{document}
\title{The 3D Flow Field Around an Embedded Planet}

\author{Jeffrey Fung \altaffilmark{1}, Pawel Artymowicz\altaffilmark{1,2} and Yanqin Wu \altaffilmark{1}}

\altaffiltext{1}{Department of Astronomy and Astrophysics, University of Toronto, 50 St. George Street, Toronto, ON, Canada M5S 3H4}
\altaffiltext{2}{Department of Physical and Environmental Sciences, University of Toronto at Scarborough, 1265 Military Trail, Scarborough, ON, Canada M1C 1A4}

\email{fung@astro.utoronto.ca}

\begin{abstract}
3D modifications to the well-studied 2D flow topology around an embedded planet have the potential to resolve long-standing problems in planet formation theory. We present a detailed analysis of the 3D isothermal flow field around a 5 Earth-mass planet on a fixed circular orbit, simulated using our multi-GPU hydrodynamics code \texttt{PEnGUIn}. We find that, overall, the horseshoe region has a columnar structure extending vertically much beyond the Hill sphere of the planet. This columnar structure is only broken for some of the widest horseshoe streamlines, along which high altitude fluid descends rapidly into the planet's Bondi sphere, performs one horseshoe turn, and exits the Bondi sphere radially in the midplane. A portion of this flow exits the horseshoe region altogether, which we refer to as the ``transient'' horseshoe flow. The flow continues as it rolls up into a pair of up-down symmetric horizontal vortex lines shed into the wake of the planet. This flow, unique to 3D, affects both planet accretion and migration. It prevents the planet from sustaining a hydrostatic atmosphere due to its intrusion into the Bondi sphere, and leads to a significant corotation torque on the planet, unanticipated by 2D analysis. In the reported simulation, starting with a $\Sigma\sim r^{-3/2}$ radial surface density profile, this torque is positive and partially cancels with the negative differential Lindblad torque, resulting in a factor of 3 slower planet migration rate. Finally, we report 3D effects can be suppressed by a sufficiently large disk viscosity, leading to results similar to 2D.
\end{abstract}

\keywords{hydrodynamics, planet-disk interactions, protoplanetary disks}

\section{Introduction}
\label{sec:intro}
The interaction between newly formed planets and the protoplanetary disks, in which they are embedded, is an integral part of planet formation theory. In recent years, the field of planetary science has made great strides with the discovery of thousands of planets and planet candidates by the \textit{Kepler} mission \citep{Kepler}. The majority of these planets have sizes of about 1 to 4 Earth-radii (super-Earths), located at about 0.1 AU away from their host stars \citep[e.g.][]{Batalha2013,Petigura2013}. This wealth of data has enabled us to more thoroughly check the accuracy of our theories against observations. 

Despite much effort in the study of planet formation, there remain some discrepancies between theory and observation that urgently need to be bridged. One example is planet migration. Through gravitational interaction with the circumstellar disk, a planet can gain or lose angular momentum and migrate away or toward its host star. Current planet migration theory predicts that Earth-size planets located at about 0.1 AU away from their host stars would migrate inward and fall onto their host stars within a timescale of only about a few thousand years \citep[][]{Masset2010,Paardekooper2010,Paardekooper2011}, 
while the typical lifetime of a protoplanetary disk is about a few million years. The fast inward migration is also known as type I migration. Because we do observe many planets at these separations from their host stars, this implies our predicted migration rate is orders of magnitude faster than can be tolerated by observational data.

Another example is the accretion of planetary atmospheres. The study of how planetary cores of a few Earth-masses ($M_{\oplus}$) accrete gas from their protoplanetary disks is essential for understanding how gaseous planets are formed. Current planet accretion theory often uses the Bondi radius $\rB$ to define the extent of an embedded planet's atmosphere \citep[e.g.][]{Pollack1996, Ikoma2000,Rafikov2006,Lee2014}.
\begin{equation}
\label{eqn:rB}
\rB = q \frac{G M_\ast}{\cs^2} \,,
\end{equation}
where $q$ is the mass ratio between the planet and the host star, $G$ is Newton's gravitational constant, $M_\ast$ is the host star's mass, and $\cs$ is the sound speed of the disk. Treatment of this form neglects the effects of the background Keplerian shear and the disk's vertical structure. On the other hand, recent results in 3D disk-planet interaction by \citet{Ormel2015b} (hereafter OSK15) found that no gas is bound to the planet over a timescale exceeding tens of planetary orbital periods. If their result applies to Earth-size planetary cores, it might prevent their transformation into gaseous giants like Neptune, since the core's envelope may be prevented from cooling, contracting, and accepting more gas from the surrounding flow. 

\begin{figure}[]
\includegraphics[width=0.99\columnwidth]{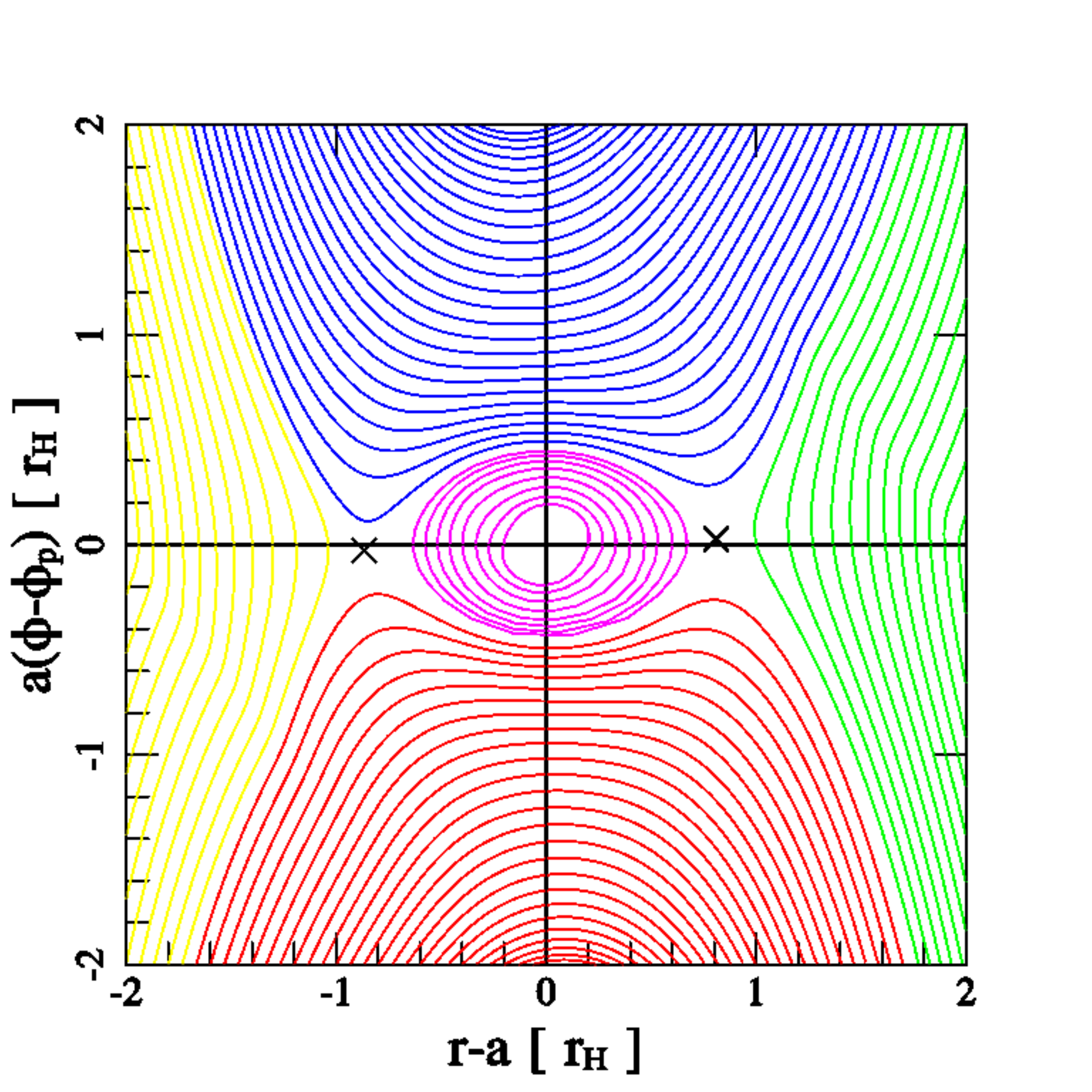}
\caption{Streamlines around a planet in 2D, plotted in the corotating frame of the planet, which is located at the center of this plot. The background Keplerian shear is from bottom to top in the inner disk ($r<a$), and top to bottom in the outer disk ($r>a$). We call the streamlines approaching the planet from the inner disk ``inner'', and those approaching from the outer disk ``outer''. The streamlines are color-coded: yellow and green are the inner and outer disk flow; red and blue are the inner and outer horseshoe flow; and magenta is the flow that is bound to the planet. The crosses mark the ``stagnation'' points, where the velocity is zero. A third stagnation point exists at the location of the planet. This point is irrelevant to our analysis, so we omit to label it. The streamlines here are computed from a 2D simulation using the same setup and resolution as our 3D one (see \secref{sec:numerics}), but without the vertical dimension, and the planet's potential is not softened.}
\label{fig:quad}
\end{figure}

The shortcomings of current theory may in part be due to our lack of understanding about the 3D flow of disk material near the gravitational influence of a planet. Both planet migration and accretion theories rely on understanding the flow topology, yet our current picture of disk-planet interaction is based on a simplified 2D flow, confined to the midplane of the disk, or even a 1D, spherically symmetric flow in the classical Bondi-Hoyle accretion. \figref{fig:quad} shows the typical 2D corotating streamlines around a planet located at $\bm{r}_{\rm p}=(a,\phi_{\rm p})$. In this picture, there are three separate, well defined flow patterns: (i) the disk flow, which can be approximated as deformed, originally circular orbits around the host star, represented by the yellow and green streamlines; (ii) corotational or horseshoe flow, which contains the horseshoe- and tadpole-shaped orbits of the restricted three-body problem, represented by the red and blue streamlines; and (iii) streamlines bound to the planet, represented by the magenta streamlines. We refer the reader to \citet{Ormel2013} for a more thorough analysis of the 2D flow topology. The following established notions in planet formation theory are tied to each of the three flow patterns listed above: (i) density waves are excited in the disk flow; the angular momentum carried by these waves results in the Lindblad torque (also called wave torque) acting on the planet; (ii) the corotational flow exchanges angular momentum with the planet as it transfers from outside to inside the planet's orbit at $r=a$, which results in the corotation torque (also called co-orbital torque or horseshoe drag); and (iii) the planet is surrounded by an atmosphere separated from the other regions of the disk.

The 2D picture described above is most applicable to large planets, whose Hill radius (or Roche lobe radius):
\begin{equation}
\label{eqn:rH}
\rH = a\left(\frac{q}{3}\right)^{\frac{1}{3}} \,,
\end{equation}
where $a$ is the semi-major axis of the planet's orbit, is larger than the local scale height of the disk, $h_0$, in order to justify the flat disk assumption in the vicinity of the planet. For super-Earths, this condition is not satisfied. For example, if the planet has $1 M_{\oplus}$, or $q\sim3\times10^{-6}$, then it has $\rH\sim0.01a$, while $h_0$ is typically between $0.03$ to $0.1a$. As a result, one cannot assume that the flow topology can be adequately described by \figref{fig:quad}. What does the horseshoe flow look like above and below the planet's Hill sphere? How does gas flow around the planet and how does it become accreted? These are the questions we aim to answer in this paper, specifically for super-Earths.

We use high-resolution global simulations to simultaneously resolve the global flow in the horseshoe region and the local flow within the planet's atmosphere. In previous works on 3D disk-planet interaction, reduction of a varying degree of migration torques has been found. Linear analysis by \citet{Tanaka2002} has given a 3D torque weaker by about a factor of two than in a 2D disk, in line with the earlier analytical estimates of \citet{Artymowicz1993}, while fully nonlinear simulations by \citet{DAngelo2003} have found that for planets with $\rH$ less than 60\% of $h_0$, the torque can be up to an order of magnitude weaker than 2D predictions; and when \citet{DAngelo2010} studied the fully nonlinear case for a small planet ($\sim 1 M_{\oplus}$), this time they have found good agreement with \citet{Tanaka2002}. Overall, it appears that 3D effects are more important for the larger than for the smaller planets, which is counter-intuitive. We will demonstrate in this paper that these results can be tied together and explained, by studying how the flow topology differs from 2D to 3D. 

In terms of the accretion of a planet's atmosphere, OSK15's results were obtained from the local 3D simulations of sub-Earth-mass planets' atmospheres, which did not allow them to self-consistently connect their atmospheric flow to the global disk flow. In this work we will not only establish how a planet's atmosphere fits into the co-orbital flow topology, but also probe the interesting regime of planet masses closer to the critical core mass limit.

\subsection{Plan of the Paper}
\label{sec:plan}
In \secref{sec:numerics} we describe our numerical setup. In \secref{sec:results} we present the flow topology extracted from our simulation. In \secref{sec:tor} we compute the torque on our planet. In \secref{sec:viscosity} we demonstrate how disk viscosity can affect the flow pattern and consequently the planetary torque. In \secref{sec:discussion} we conclude and discuss the implications of our results.

\section{Setup}
\label{sec:numerics}
To perform global 3D hydrodynamical simulations of disk-planet interaction, we use the Lagrangian, dimensionally-split, shock-capturing hydrodynamics code \texttt{PEnGUIn} ({\bf P}iecewise Parabolic Hydro-code {\bf En}hanced with {\bf G}raphics Processing {\bf U}nit {\bf I}mplementatio{\bf n}), which has previously been used to simulate disk gaps opened by massive planets \citep{Fung2014}. This code is written in CUDA-C, and runs on multiple GPUs (Graphics Processing Units). \texttt{PEnGUIn} uses the Lagrangian-remap formulism of the piecewise parabolic method (PPM; \citealt{PPM}), which is a high-order Godunov scheme, with improvements suggested by \citet{VH1} included.


Our simulations are done on a fixed \footnote{While \texttt{PEnGUIn} solves hydrodynamics equations in the Lagrangian frame, the solutions are remapped onto a fixed grid after every timestep. See \citet{PPM} for details.} cylindrical grid. It spans the full $2\pi$ in the azimuth, $0.7a$ to $1.3a$ in the radial direction, and $0a$ to $0.15a$ in the vertical, where $a$ is the fixed planet-star separation. \texttt{PEnGUIn} solves the following Navier-Stokes equations in cylindrical coordinate:

\begin{equation}
\label{eqn:cont_eqn}
\frac{{\rm D}\rho}{{\rm D} t} = -\rho\left(\bm{\nabla}\cdot\bm{v}\right) \,,
\end{equation}
\begin{equation}
\label{eqn:moment_eqn}
\frac{{\rm D} \bm{v}}{{\rm D} t} = -\frac{1}{\rho}\bm{\nabla} p - \bm{\nabla} \Phi + \frac{1}{\rho}\bm{\nabla}\cdot\mathbb{T}  \,,
\end{equation}
where ${\rm D}/{\rm D}t$ is the Lagrangian derivative, $\rho$ is the density, $p$ the gas pressure, $\bm{v}$ the velocity, $\mathbb{T}$ the Newtonian viscous stress tensor which depends on the kinematic viscosity $\nu$, and $\Phi$ the gravitational potential of the central star and the planet. In the center-of-mass frame,
\begin{align}
\label{eqn:potential}
\Phi = &-\frac{GM_\ast}{\sqrt{r^2 + r_{\rm 1}^2 + 2 r r_{\rm 1}\cos(\phi-\phi_{\rm p}) + z^2}} \nonumber \\
&-\frac{qGM_\ast}{\sqrt{r^2 + r_{\rm 2}^2 - 2 r r_{\rm 2}\cos(\phi-\phi_{\rm p}) + z^2 + \rs^2}} \, ,
\end{align}
where $r_{1} = q a/(1+q)$ and $r_{2}=a/(1+q)$ are the star's and the planet's radial positions, respectively; $\phi_{\rm p}-\pi$ and $\phi_{\rm p}$ are their angular positions; and $\rs$ is the smoothing (a.k.a.~softening) length of the planet's potential. Recall that $M_\ast$ is the mass of the star and $q$ is the mass ratio between the planet and the star. We set $GM_\ast(1+q)=1$ and $a=1$ so that the planet's orbital frequency $\Omega_{\rm p}=1$ and period $P_{\rm p}=2\pi$. For convenience, we also denote $v_{\rm k}=\sqrt{GM_\ast(1+q)/r}$ and $\Omega_{\rm k}=\sqrt{GM_\ast(1+q)/r^3}$ as the Keplerian orbital velocity and frequency. We complete our set of equations with an isothermal equation of state: $p=\cs^2 \rho$, where $\cs$ is the constant sound speed of the disk.

Like in \citet{Fung2014}, \texttt{PEnGUIn} performs in the planet's co-rotating frame, but uses the conservative form of the angular momentum equation in place of explicitly computing the Coriolis force \citep{Kley98}. Also, we note that in its most basic form, Godunov scheme suffers inaccuracy in resolving hydrostatic equilibrium, because the classical Riemann problem is constructed without consideration to the source term, which modifies the momentum of the fluid only after the flux of conservative quantities has been calculated. This typically results in spurious oscillations in the numerical solutions. PPM accounts for this by embedding the source term in the Riemann solver, as shown by Equations 2.9 in \citet{PPM}. We have tested and verified that \texttt{PEnGUIn} adequately maintains a 3D disk in hydrostatic equilibrium in the absence of a planet.

\subsection{Planet and Disk Parameters}
\label{sec:parameters}
We simulate a planet on a fixed circular orbit embedded in a viscous 3D disk. The planet-to-star mass ratio is $q=1.5\times 10^{-5}$, which corresponds to $\sim 5M_{\oplus}$ for a solar-mass star. We increase the planet mass gradually at the beginning of our simulations, starting from zero to our desired value over $1 P_{\rm p}$. The relevant length scales in this study are $h_0$, the scale height of the disk, $\rH$, the Hill radius within which the gravity of the planet dominates, $\rB$, the Bondi radius relevant for accretion, and $\rs$, the smoothing length. Since $\cs$ is constant, we have $h=\cs/\Omega_{\rm k}$, which has a radial dependence that goes as $h=h_0 (r/a)^{\frac{3}{2}}$, and is equal to $h_0=0.03a$ at $a$. Our planet has $\rH=0.017a$ (\eqnref{eqn:rH}), and $\rB=qGM_\ast/\cs^2=0.0167a$. Finally, for $\rs$, unlike in 2D calculations where a non-zero $\rs$ mainly serves as a way to mimic 3D effects, here we include it to avoid singularity. We choose $\rs=0.1\rH$, or $0.0017a$. Our set of parameters therefore gives us the following hierarchy of length scales: $h>\rH\sim\rB>\rs$.

Another convenient way to quantify the planet's mass is the dimensionless ``thermal mass'':
\begin{equation}
q_{\rm th} = q\left(\frac{h_0}{a}\right)^{-3} \, ,
\label{eqn:q_th}
\end{equation}
which can also be written as $\rB/h_0$. The value $q_{\rm th}\approx 1$ marks the division line where a planet becomes sufficiently massive to significantly modify the disk structure, which is sometimes called the ``nonlinear'' regime. Because we have $q_{\rm th}=0.56$, which is close to unity, we do expect our planet to have some weak nonlinear effects. The shortest length scale $\rs$ can be interpreted as the physical size of the planet. The Earth-Sun system, for example, would have $\rs\sim0.04\rH$ if the Earth were placed at $0.1$AU. Since $\rs$ is much smaller than both $\rH$ and $\rB$, we expect it to have little influence on the global flow topology, but has more influence on the density profile of the planet's atmosphere.

The initial density profile of the disk is axisymmetric. It follows a power law in the radial direction, and is set to the hydrostatic solution of an isothermal gas in the vertical direction:
\begin{equation}
\rho(r,z) = \rho_0 \left( \frac{r}{a} \right)^{-3} e^{-\frac{z^2}{2h^2}} \, ,
\label{eqn:rho}
\end{equation}
where we set $\rho_0=1$\footnote{Since we do not consider the self-gravity of the disk, this normalization has no impact on our results.}. We impose this initial profile everywhere in our simulation domain, including the vicinity of where the planet will gradually appear. The surface density, $\Sigma$, obtained by integrating \eqnref{eqn:rho} over $z$, has the following profile: $\Sigma\propto r^{-\frac{3}{2}}$. This surface density profile is intentionally chosen to test a prediction about corotation torque. It has been shown in 2D that the corotation torque vanishes when there is zero disk vortensity gradient, or $(\bm{\nabla}\times\bm{v})/\Sigma =$ constant \citep{Ward1991,Masset2004,Paardekooper2010}, which, for a Keplerian disk, is precisely when $\Sigma\propto r^{-\frac{3}{2}}$. Consequently, if we find a significant corotation torque in our 3D simulation, it will be a new phenomenon not captured by 2D analysis.

The initial velocity field models a steady disk by setting $\partial/\partial t = 0$ in \eqnref{eqn:moment_eqn} and ignoring the potential of the planet: ${\bm v}=(v_{\rm r}, r\Omega, 0)$, where
\begin{equation}
v_{\rm r} = -3\frac{\nu}{r}\left(\ddlnr{\ln{\rho}}+\frac{1}{2}\right) \, ,
\label{eqn:vr}
\end{equation}
\begin{equation}
\Omega = \Omega_{\rm k}\sqrt{ 1+ \left(\frac{h}{r}\right)^2 \ddlnr{\ln{\rho}} } \, ,
\label{eqn:freq}
\end{equation}

The kinematic viscosity of our disk is set to $\nu=10^{-6}a^2\Omega_{\rm p}$, which corresponds to the Shakura-Sunyaev $\alpha$-viscosity coefficient $\alpha\sim0.001$ \citep{alpha}. This choice determines the viscous diffusion timescale across the horseshoe region. One can estimate this timescale as $t_{\nu}=w^2/\nu$, where $w$ is the half of the radial width of the horseshoe region. Viscous diffusion can modify the horseshoe flow if $t_{\nu}$ is shorter than the libration timescale of the widest horseshoe orbit, $t_{\rm lib}=(4a/3w)P_{\rm p}$. If we approximate $w\sim 2\rH$, then our model gives $t_{\nu}\sim 200P_{\rm p}$ which is much longer than $t_{\rm lib}\sim 40P_{\rm p}$. As a result, it is safe to neglect the effects of viscosity in our analysis of the flow topology. In \secref{sec:cor_flow} we will give an exact measurement of $w$, and in \secref{sec:viscosity} we will consider the scenario where $t_{\nu}<t_{\rm lib}$.

\subsection{Boundary Conditions.}
\label{sec:grid}
Because we cover the full $2\pi$ span, our azimuthal boundary condition is periodic. For our radial boundaries, it is important to prevent the reflection of density waves launched by the planet. To achieve this, we impose wave killing zones in $r=\{0.7a,0.73a\}$ for the inner boundary and $r=\{1.27a,1.3a\}$ for the outer. Within these zones, we include an artificial damping term:
\begin{equation}
\frac{\partial{X}}{\partial t} = \left[X(t=0)-X(t)\right]\frac{2\cs|r-r_{\rm kill}|}{(0.03a)^2} \, ,
\end{equation}
where $X$ stands for all disk variables including $\rho$, $p$, and $\bm{v}$. $r_{\rm kill} = 0.73a$ for the inner boundary, and $1.27a$ for the outer. In the vertical direction, as we only simulate the upper half the of the disk, so we impose a symmetric condition at the disk midplane. At the top we use a reflecting condition to ensure that mass does not leak in or out. In practice, the symmetric and reflecting conditions are equivalent: we copy disk variables across the boundary, and reverse the signs of the $z$-component velocity and force.

\subsection{Resolution}
\label{sec:resolution}
\begin{figure}[]
\includegraphics[width=0.99\columnwidth]{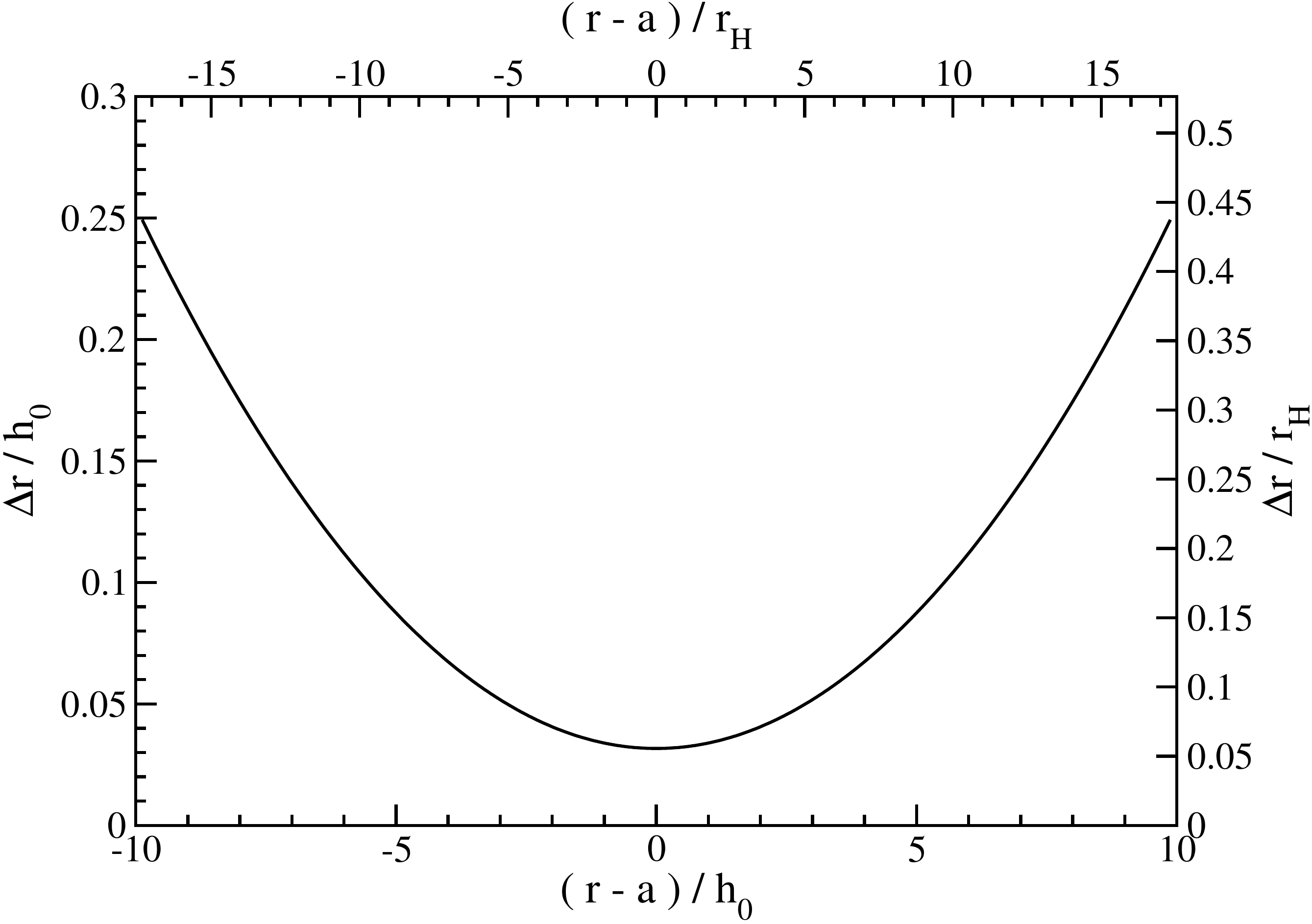}
\caption{Radial resolution of our grid as described by Equations \ref{eqn:res} and \ref{eqn:resA}. Near the planet's location, we have $\sim 32$ cells per $h_0$, or $\sim 18$ cells per $\rH$.}
\label{fig:res}
\end{figure}

Since the focus of this work is to study the co-orbital region, we adopt a nonuniform grid that has enhanced resolution close to the planet. Individual grid size is calculated using the following formula:
\begin{equation}
\Delta x = A (x-x_{\rm p})^2 + \Delta x_{\rm min} \, ,
\label{eqn:res}
\end{equation}
where $x$ can be any one of $r,\phi,$ or $z$, $\Delta x$ is the grid size, $x_{\rm p}$ is the location of the planet, and $\Delta x_{\rm min}$ is the desired grid size at the location of the planet. The factor $A$ is found using the following relation:
\begin{equation}
x_{\rm b} = \sqrt{\frac{\Delta x_{\rm min}}{A}} \tan (NA\Delta x_{\rm min}) \, ,
\label{eqn:resA}
\end{equation}
where $x_{\rm b}$ is the distance from the planet's location to the boundary of the grid, which equals to $0.3$ in $r$, $\pi$ in $\phi$, or $0.15$ in $z$ . $N$ is the total number of grid cells within $x_{\rm b}$, which equals to $144$ in $r$, $1512$ in $\phi$, or $72$ in $z$. The entire grid therefore has $288(r)\times3024(\phi)\times72(z)$ cells. $\Delta x_{\rm min}$ defines the resolution we desire near the planet, which we choose to be $\rH/18\sim 0.001a$. This also corresponds to about 2 cells per $\rs$. \figref{fig:res} illustrates how the radial cell size changes across the grid. This prescription has the advantage of creating a high resolution region near the planet that has a roughly uniform cell size, without introducing abrupt changes in resolution which are prone to generating numerical error. \figref{fig:xs} demonstrates the level of convergence our resolution has achieved with respect to our measurement of the horseshoe half-width, $w$. Reducing our resolution by a factor of 1.2, or even 1.44, changes our answer by about $1\%$.

At this resolution, we have $\sim 63$ million cells in total. Using 3 GTX-Titan graphics cards housed in a single desktop computer, our GPU-accelerated code \texttt{PEnGUIn} runs at a speed of 0.84 seconds per timestep, or about 140 minutes per planetary orbit.

\section{Flow Topology}
\label{sec:results}
We run our simulations for $100P_{\rm p}$ to reach a steady state, and then compute the time-averaged density and velocity field from $100$ to $101P_{\rm p}$ to obtain our final results. \secref{sec:cor_flow} will describe the overall size and shape of the horseshoe region; \secref{sec:ver_flow} will focus on the horseshoe turn, where a close encounter with the planet occurs, and show how this flow interacts with the planet's atmosphere; and finally, \secref{sec:tur_flow} will investigate the aftermath of the close encounter.

\subsection{The Horseshoe Region}
\label{sec:cor_flow}
Near the midplane, we expect to find horseshoe orbits, like those in \figref{fig:quad}. If we go to a higher altitude, above the planet's Hill sphere, it becomes unclear what kind of trajectory the gas will take to flow around, or across, the Hill sphere. One simple question we can ask is how far up in altitude does the horseshoe region extend.

We use our velocity data to reconstruct the fluid streamlines. At a given $z$, the largest $|\Delta r| = r-a$ that still performs a horseshoe turn is defined as $w$, the horseshoe half-width at that $z$; we will also refer to this streamline as the ``widest horseshoe orbit''. We measure $w$ at $\phi=\phi_{\rm p}-1$ for the inner flow, and $\phi_{\rm p}+1$ for the outer one\footnote{It is also possible to measure $w$ post-horseshoe turn, by tracking the streamlines backward in time, but we find the flow there too complex for a clean measurement. See \secref{sec:tur_flow} for a discussion on the flow topology there.}. We find the two differ by about $\sim1\%$, with the inner orbits being the wider one. We consider this difference insignificant for the scope of this paper. The $w$ we report is the average of the two.

\begin{figure}[]
\includegraphics[width=0.99\columnwidth]{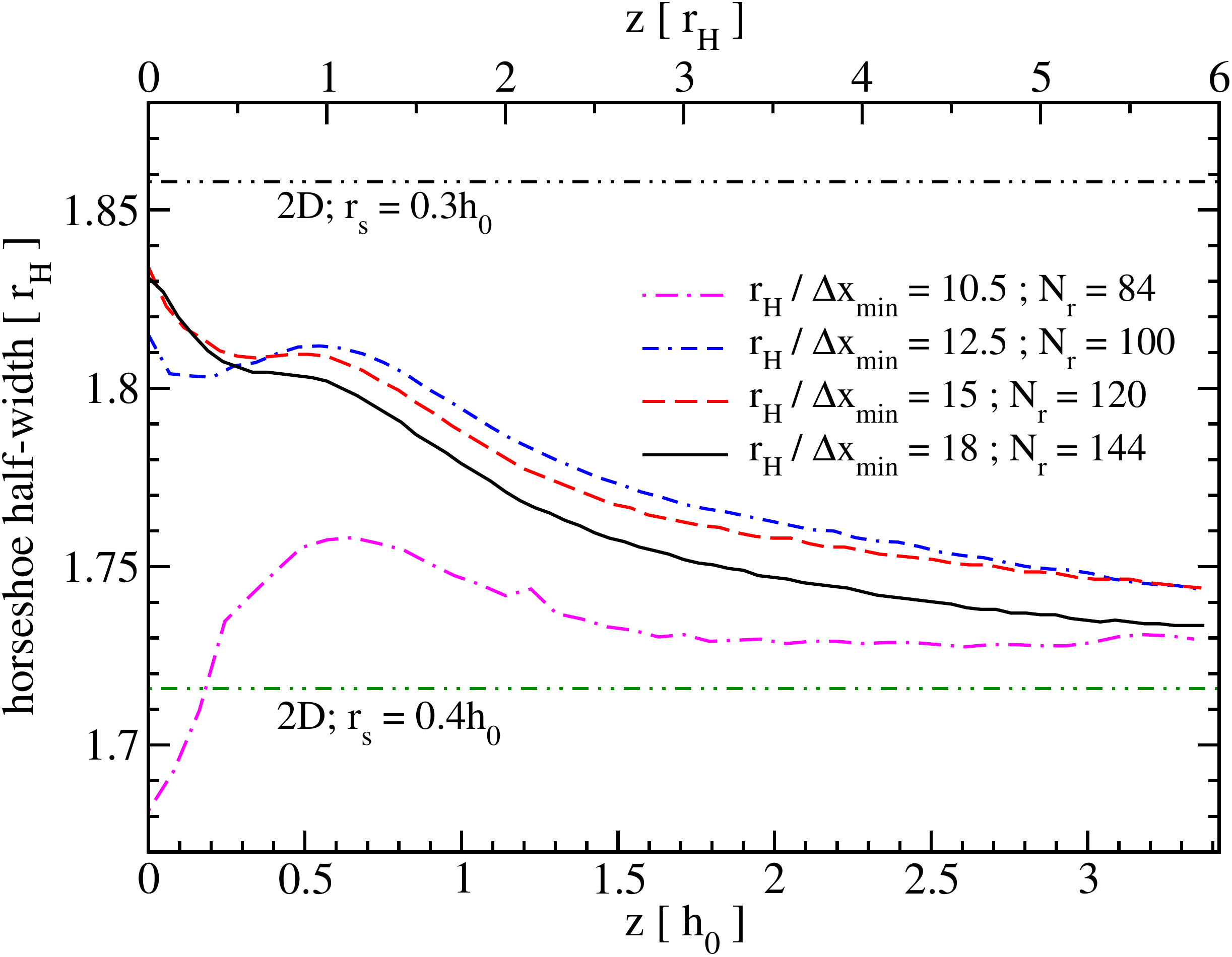}
\caption{Horseshoe half-width as a function of height above the midplane. $z$ refers to the height of the flow before the turn. The magenta dot-dashed curve, blue dot-dash-dashed curve, red dashed curve, and black solid curve are results from different simulations, where the resolution is 20\% higher between each curve. The black solid curve is our choice of resolution. This plot shows that our measurement has converged to within 1\%. Also shown for comparison are results from 2D simulations with different smoothing lengths, at the same resolution as the black solid curve (also see \figref{fig:xs_2D}).}
\label{fig:xs}
\end{figure}

\begin{figure}[]
\includegraphics[width=0.99\columnwidth]{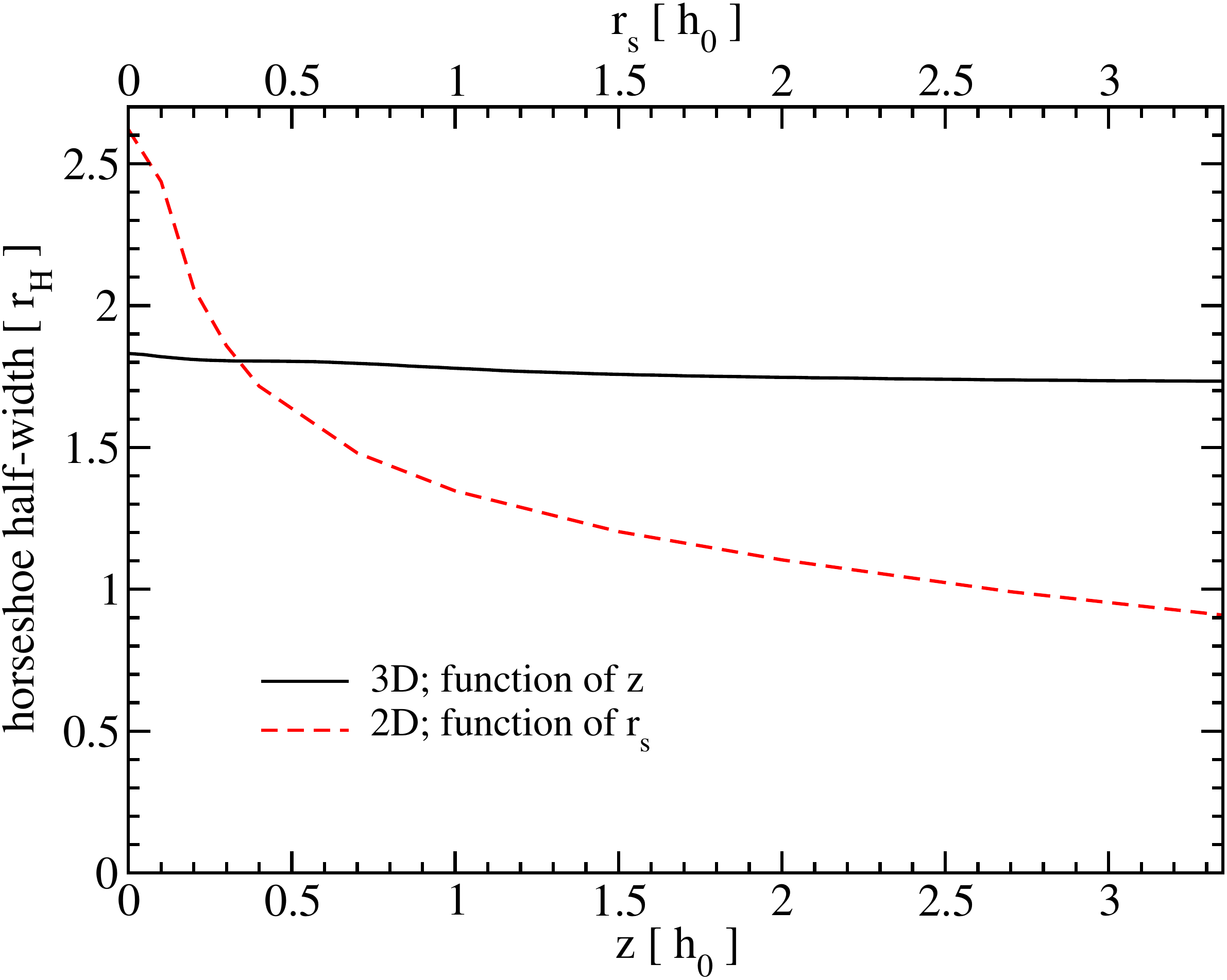}
\caption{Horseshoe half-width as a function of height above the midplane in 3D, and a function of smoothing length in 2D. The black solid curve is the same as the black curve in \figref{fig:xs}. The red dashed curve is from a series of 2D simulations. This plot demonstrates that a 3D disk behaves differently from a combination of 2D layers.}
\label{fig:xs_2D}
\end{figure}

\figref{fig:xs} plots $w$ as a function of $z$. Remarkably, $w$ remains nearly constant in $z$, even when $z$ reaches $6\rH$ or $\sim3h_0$. Its average value, weighted by disk density, is $\sim1.8\rH$, and the corresponding libration time for the widest horseshoe orbit is $t_{\rm lib}\approx 43P_{\rm p}$. This width is wider than what is expect for planets in the linear regime ($q_{\rm th}\ll 1$), which is $w\sim1.2~a\sqrt{aq/h_0}$ \citep{Masset2006, Paardekooper2009}. In these units, our planet has $w=1.35~a\sqrt{aq/h_0}$. This is consistent with the findings of \citet{Masset2006}, where they showed that as $q_{\rm th}$ increases to unity, there is an increase in the horseshoe half-width compared to linearly estimated values. We also perform a series of 2D simulations, with varying smoothing lengths, to compare with our result. In \figref{fig:xs_2D}, we see that if one stacks layers of 2D disks, increasing $\rs$ to mimic the effect of increasing altitude, one will not recover the same horseshoe half-width we find in 3D. 

\begin{figure*}[]
\includegraphics[width=1.99\columnwidth]{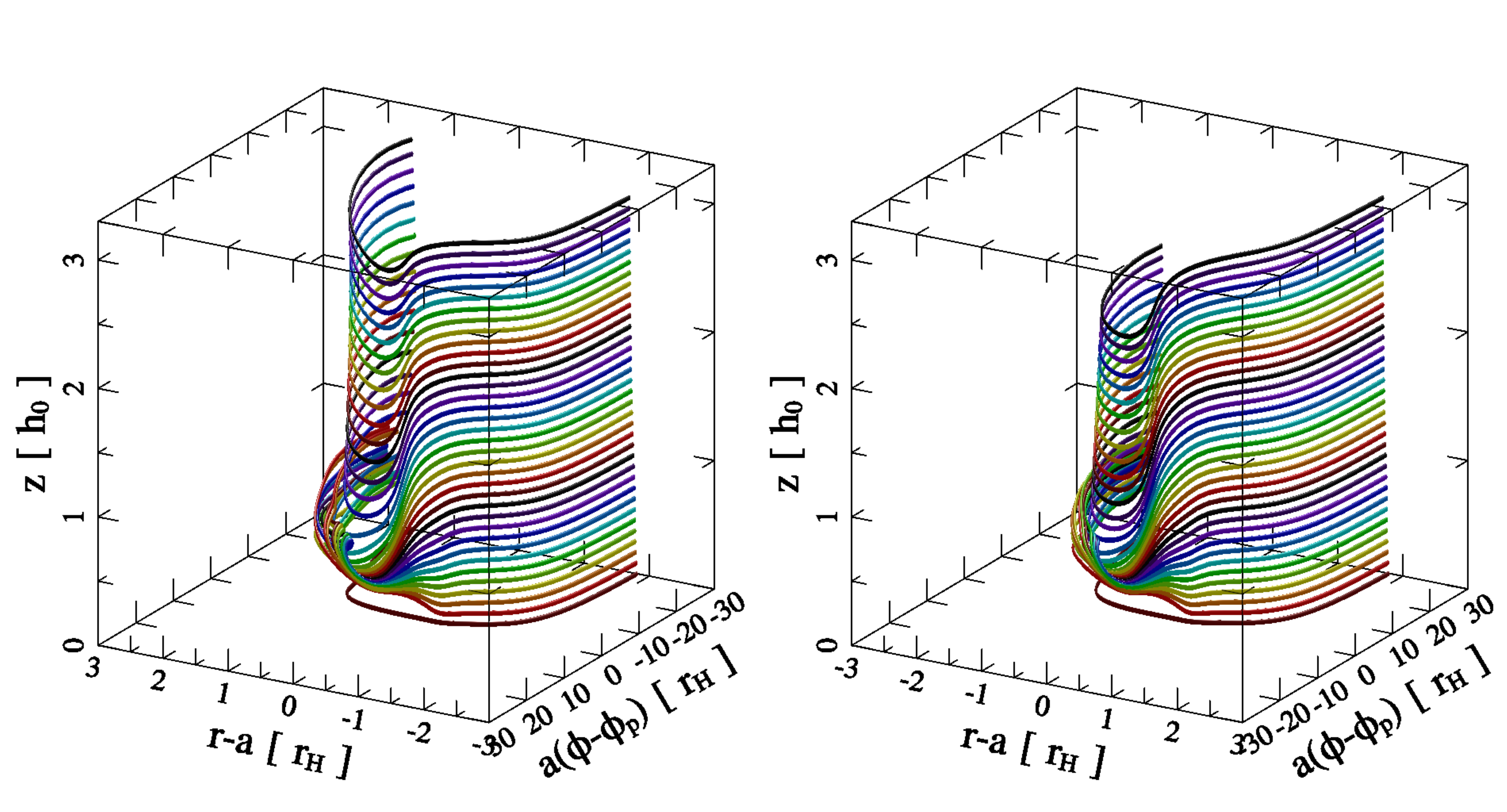}
\caption{Streamlines of the widest horseshoe orbits. The left panel shows the inner flow, and the right shows the outer one. Note that 1) the flow has a columnar structure along the horseshoe turn; 2) most streamlines go through a sharp drop in altitude half-way through their turns, being drawn vertically to the planet; 
3) a more complex flow structure is seen near the midplane after the turn; see \figref{fig:mix} for a close-up picture of the streamlines there. }
\label{fig:hs}
\end{figure*}

To help visualize these trajectories, \figref{fig:hs} plots the 3D streamlines of the widest horseshoe orbits. Clearly the horseshoe region has a columnar structure. It additionally shows that halfway through the widest horseshoe turns, as the flow crosses the planet's orbit at $r=a$, some of it rapidly accelerates vertically toward the planet, and completes the turns at a significantly lower altitude. In fact, all fluid elements that started at $z\leq 2h_0$ are pulled down to $z\lesssim\rH$ after their horseshoe turns.

On one hand, this flow is columnar; there is almost no vertical variation in the planar velocities. This appears akin to Taylor-Proudman columns, where $\partial\bm{v}/\partial z\approx 0$ due to a dominating Coriolis force. On the other hand, the vertical flow directly above (and below) the planet implies a rapid $v_{\rm z}$ variation in $z$, and a more complex, non-columnar flow structure can be seen immediately after the horseshoe turn (\figref{fig:hs}); these flow patterns do not follow Taylor-Proudman theorem. In the following section, we will show analytically that a columnar flow structure is expected of a Keplerian disk; at the same time, we will account for the non-columnar features we observe.

\subsubsection{Columnar Flow in a Keplerian Disk}
\label{sec:theory}
We rewrite \eqnref{eqn:moment_eqn} in a rotating frame (such as the rotating frame of a planet), where $\bm{u}$ is the velocity in the rotating frame and $\bm{\Omega}=(0,0,\Omega_{\rm p})$ is the frame's rotation frequency.
\begin{equation}
\frac{\partial \bm{u}}{\partial t} + (\bm{u}\cdot\bm{\nabla})\bm{u} + 2\bm{\Omega}\times\bm{u} + \bm{\Omega}\times(\bm{\Omega}\times\bm{r}) = -\bm{\nabla}\Phi - \frac{1}{\rho}\bm{\nabla} p+\frac{1}{\rho}\bm{\nabla}\cdot\mathbb{T} \, ,
\label{eqn:momentum}
\end{equation}
where the terms on the left-hand side (LHS) are the time-dependent term, the advection term, the Coriolis force term, and the centrifugal force term; on the right-hand side (RHS) are gravity represented by potential $\Phi$, pressure gradient force, and the viscous stress.  For a barotropic, steady flow with a large Reynolds number ($Re\approx |\bm{u}|h_0/\nu\gg 1$), we can transform \eqnref{eqn:momentum} into the vorticity equation by taking the curl of both sides,
\begin{equation}
(\bm{u}\cdot\bm{\nabla})\bm{\omega} = (\bm{\omega}\cdot\bm{\nabla})\bm{u} - \bm{\omega}(\bm{\nabla}\cdot\bm{u}) \, ,
\label{eqn:vorticity}
\end{equation}
where $\bm{\omega}=\bm{\nabla}\times\bm{u}+2\bm{\Omega}$ is the total vorticity. The second term on the RHS contains $\bm{\nabla}\cdot\bm{u}$, which describes the compressibility of the fluid. We can eliminate this term by combining \eqnref{eqn:vorticity} with \eqnref{eqn:cont_eqn} to obtain:
\begin{equation}
(\bm{u}\cdot\bm{\nabla})\bm{\xi} = \left(\bm{\xi}\cdot\bm{\nabla}\right)\bm{u} \, .
\label{eqn:vortensity}
\end{equation}
where
\begin{equation}
\bm{\xi} = \frac{\bm{\omega}}{\rho} = \frac{\bm{\nabla}\times\bm{u}+2\bm{\Omega}}{\rho} \,
\label{eqn:vortensity_def}
\end{equation}
is the vortensity (or potential vorticity) of the fluid. On the LHS of \eqnref{eqn:vortensity} we have the advection of $\bm{\xi}$, and on the right is the vortex tilting term. Consider an incompressible flow ($\rho =$ constant), and the case of a small perturbation in vorticity ($|2\bm{\Omega}|\gg|\bm{\nabla}\times\bm{u}|$). This gives $\bm{\xi}\approx2\bm{\Omega}/\rho=$ constant. Additionally, because $\bm{\Omega}$ is non-zero only in the $z$ direction, the RHS also simplifies, and \eqnref{eqn:vortensity} is reduced to $\partial\bm{u} /\partial z = 0$, which is the classical Taylor-Proudman theorem stating that there is no vertical variation in the flow.

For a planet's horseshoe region, we can apply a similar analysis, but with relaxed assumptions. First, instead of the incompressibility assumption, we allow the fluid to be compressible, but restrict it to a subsonic flow. Quantitatively this means the shortest length scale over which $\rho$ is allowed to vary is $h$. The local Keplerian shear is supersonic far away from the planet, so this assumption also restricts us to a radial range of $r\sim a\pm h$; however, in practice, the Keplerian shear does not usually generate shocks in the disk, so as long as $\rho$ is smooth, our analysis can apply to a larger radial range. Second, we note that a Keplerian disk does not satisfy the assumption $|2\bm{\Omega}|\gg|\bm{\nabla}\times\bm{u}|$, since $|\bm{\nabla}\times\bm{u}|=3\Omega_{\rm k}/2$; instead, we assume the vorticity of the disk is mainly in the $z$ direction, such that $\omega_z\gg\omega_r,~\omega_\phi$. By our assumptions, the LHS of \eqnref{eqn:vortensity} has a magnitude less than $\frac{\cs}{h}\left|\bm{\xi}\right|$. Therefore in component form, \eqnref{eqn:vortensity} can be estimated as:
\begin{align}
\label{eqn:newTP11}
\frac{\cs}{h}\left|\omega_r \right| \gtrsim & \left|\omega_r\frac{\partial u_r}{\partial r} + \omega_\phi\frac{1}{r}\frac{\partial u_r}{\partial \phi} + \omega_z\frac{\partial u_r}{\partial z}\right| \, ,\\
\label{eqn:newTP12}
\frac{\cs}{h}\left|\omega_{\phi} \right| \gtrsim & \left|\omega_r\frac{\partial u_\phi}{\partial r} + \omega_\phi\frac{1}{r}\frac{\partial u_\phi}{\partial \phi} + \omega_z\frac{\partial u_\phi}{\partial z}\right| \, ,\\
\label{eqn:newTP13}
\frac{\cs}{h}\left|\omega_z \right| \gtrsim & \left|\omega_r\frac{\partial u_z}{\partial r} + \omega_\phi\frac{1}{r}\frac{\partial u_z}{\partial \phi} + \omega_z\frac{\partial u_z}{\partial z}\right| \, .
\end{align}
Since the flow is subsonic, we can further approximate $\frac{\cs}{h}\gtrsim \left|\frac{\partial \bm{u}}{\partial r}\right| , \frac{1}{r}\left|\frac{\partial \bm{u}}{\partial \phi}\right|$; note that $\left|\frac{\partial (v_{\rm k}-r\Omega_{\rm p})}{\partial r}\right|\sim \frac{3}{2}\Omega_{\rm p} = \frac{3}{2}\frac{\cs}{h}$, which is within an order of unity to our approximation. Finally, rearranging \eqnref{eqn:newTP11} to \ref{eqn:newTP13}, our order-of-magnitude analysis gives:
\begin{align}
\label{eqn:newTP21}
\left|\frac{\partial u_r}{\partial z}\right| \lesssim & \frac{\cs}{h}  \left(2\left|\frac{\omega_r}{\omega_z}\right| + \left|\frac{\omega_\phi}{\omega_z}\right| \right) \, ,\\
\label{eqn:newTP22}
\left|\frac{\partial u_\phi}{\partial z}\right| \lesssim & \frac{\cs}{h} \left(\left|\frac{\omega_r}{\omega_z}\right| + 2\left|\frac{\omega_\phi}{\omega_z}\right| \right) \, ,\\
\label{eqn:newTP23}
\left|\frac{\partial u_z}{\partial z}\right| \lesssim & \frac{\cs}{h} \, .
\end{align}
This demonstrates that in the planet's co-orbital region, the variation of $v_{\rm r}$ and $v_{\phi}$ in the $z$ direction is suppressed by a factor of ${\rm max}\left(\left|\frac{\omega_r}{\omega_z}\right|, \left|\frac{\omega_\phi}{\omega_z}\right| \right)$; on the other hand, the vertical motion of the gas is allowed to vary over a length scale as short as $h$.

The physical interpretation of this result can be found in \eqnref{eqn:vortensity}, which states that a vortex line can only be tilted as much as advection can carry. If there is little planar vorticity to begin with, the advection of vorticity will not be strong enough to tilt the vortex lines to the in the $r$ and $\phi$ directions, and as a result the flow must remain columnar. $v_{\rm z}$, on the other hand, is aligned with the vortex lines, so a vertical acceleration will only lead to the compression or stretching of the vortex lines, which is allowed through the compressibility of the fluid.

\begin{figure}[]
\includegraphics[width=0.99\columnwidth]{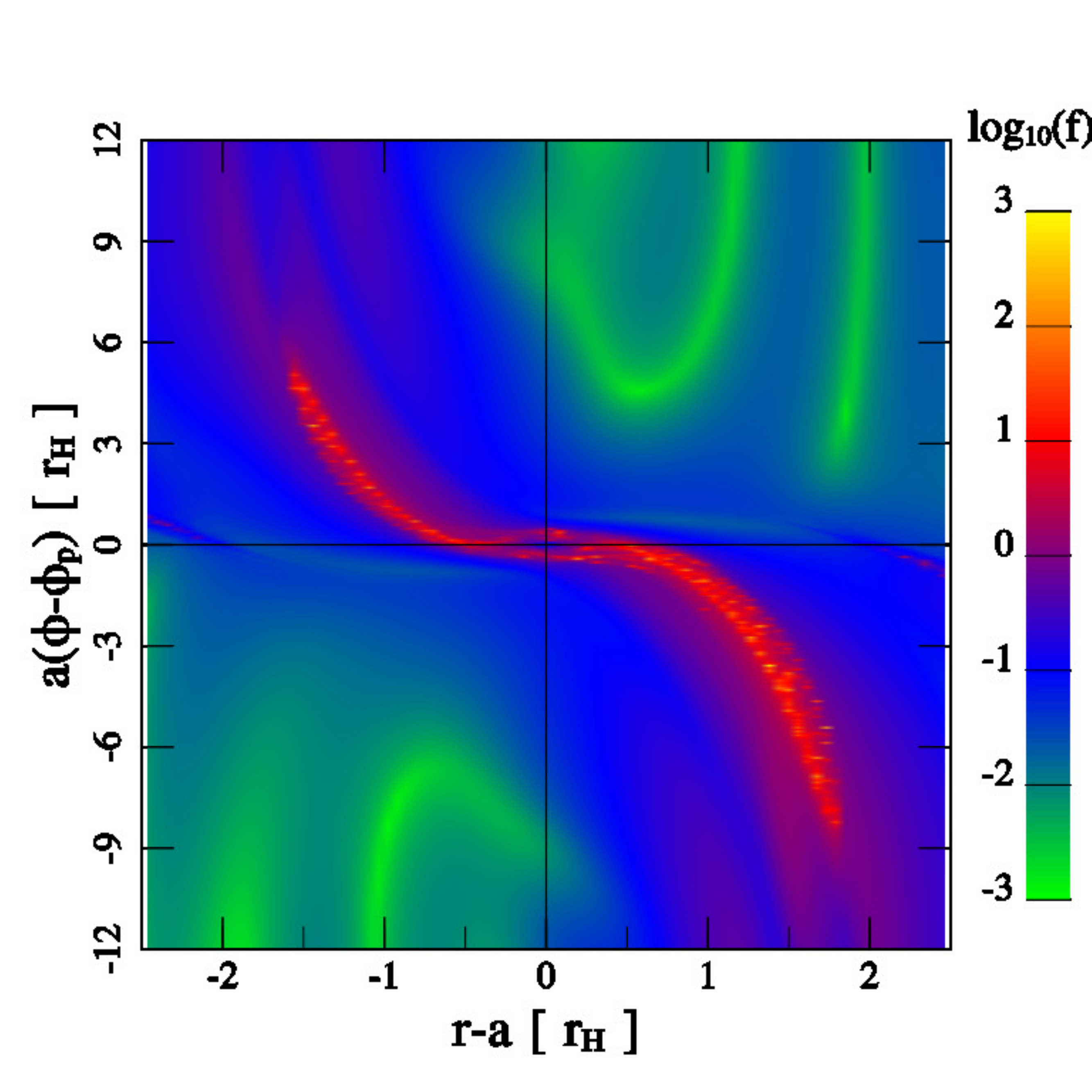}
\caption{Density-weighted vertical average of the planar-to-$z$ vorticity ratio (see \eqnref{eqn:fvor}), plotted as a function of $r$ and $\phi$. Note that $f\ll1$ in most regions, except for two streams corresponding to the finishing half of the widest inner and outer horseshoe turns.}
\label{fig:fvor}
\end{figure}

We now verify whether the assumption $\omega_z\gg\omega_r,~\omega_\phi$ holds in the flow around an embedded planet. In \figref{fig:fvor}, we plot the ratio $f$:
\begin{equation}
f\equiv \frac{ \int_{0}^{\infty} \rho \frac{\sqrt{\omega_{\rm r}^2 + \omega_{\phi}^2}}{|\omega_{\rm z}|}~ {\rm d}z}{ \int_{0}^{\infty} \rho ~{\rm d}z} \, .
\label{eqn:fvor}
\end{equation}
We find $f\ll1$ in two regions: one where $r>a$ and $\phi>\phi_{\rm p}$, corresponding to the starting half of the outer horseshoe orbits before crossing the planet's orbit; and another one where $r<a$ and $\phi<\phi_{\rm p}$, which is the starting half of the inner horseshoe flow. This is consistent with where columnar structure is found (see \figref{fig:hs}). In the regions corresponding to the finishing half of the widest inner and outer horseshoe turns, we find $f\gg1$, indicating the planar vorticity overtakes vorticity in $z$. This corresponds to the more complex flow structure after the horseshoe flow crosses $r=a$ near the midplane (see \figref{fig:hs}). In \secref{sec:tur_flow} we will further investigate this aspect of the flow.

\subsection{Flow in the Planet's Bondi Sphere}
\label{sec:ver_flow}

\begin{figure}[]
\includegraphics[width=0.99\columnwidth]{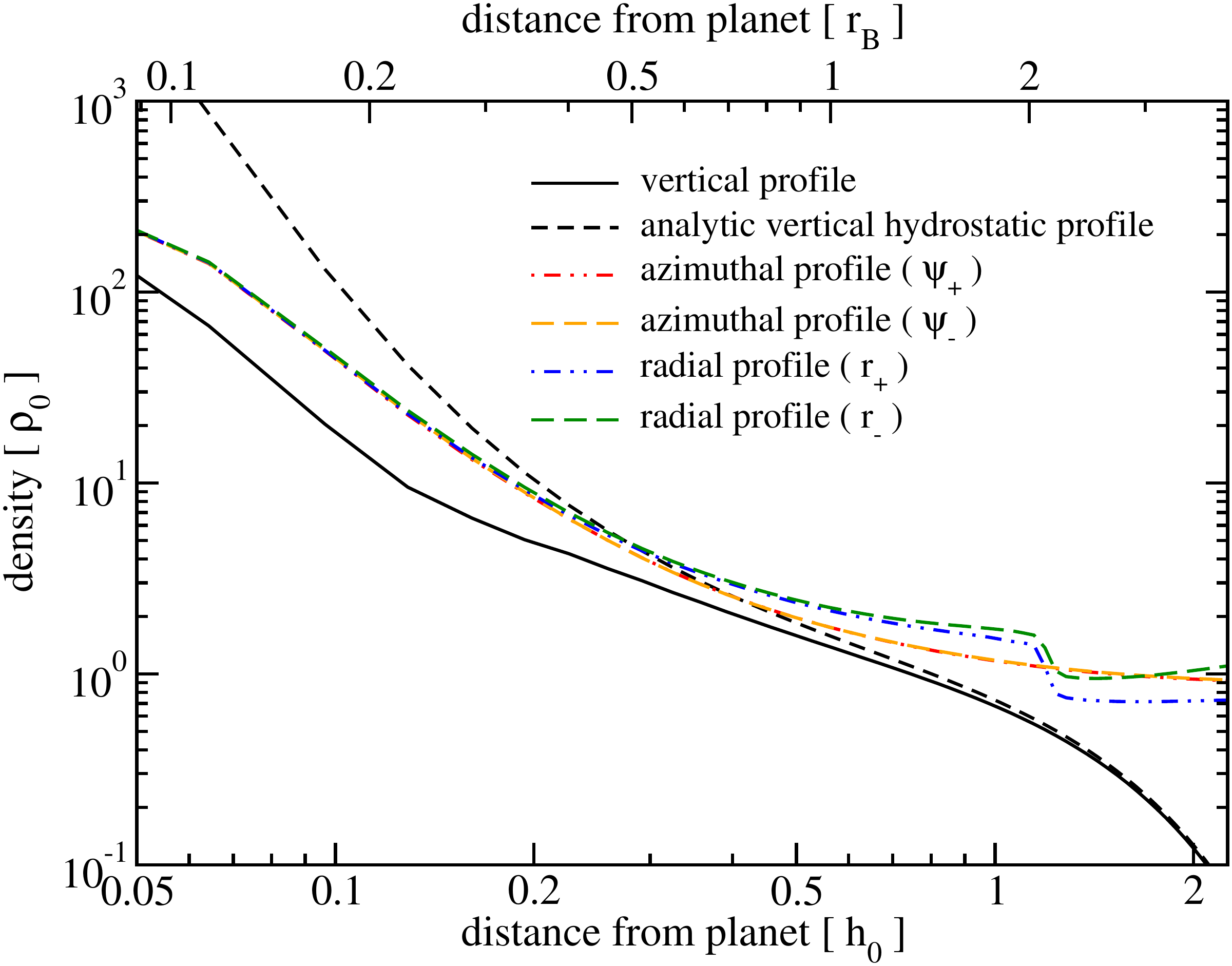}
\caption{Gas density as a function of distance from the planet. Black solid curve plots the vertical density profile; red dash-dot-dotted and orange dashed curves are both azimuthal profiles in the midplane, but in the increasing and decreasing direction of $\phi$ respectively; similarly, blue dash-dot-dotted and green dashed curves plot the radial profiles in the midplane, and are in the increasing and decreasing direction of $r$. Black dashed curve is calculated with \eqnref{eqn:static}. Note the large discrepancy between the black solid and black dashed curves. Comparing the black solid curve to the four profiles in the midplane, we see that the density structure near the planet is flattened by a factor of about $2/3$.}
\label{fig:atm}
\end{figure}

\begin{figure}[]
\includegraphics[width=0.99\columnwidth]{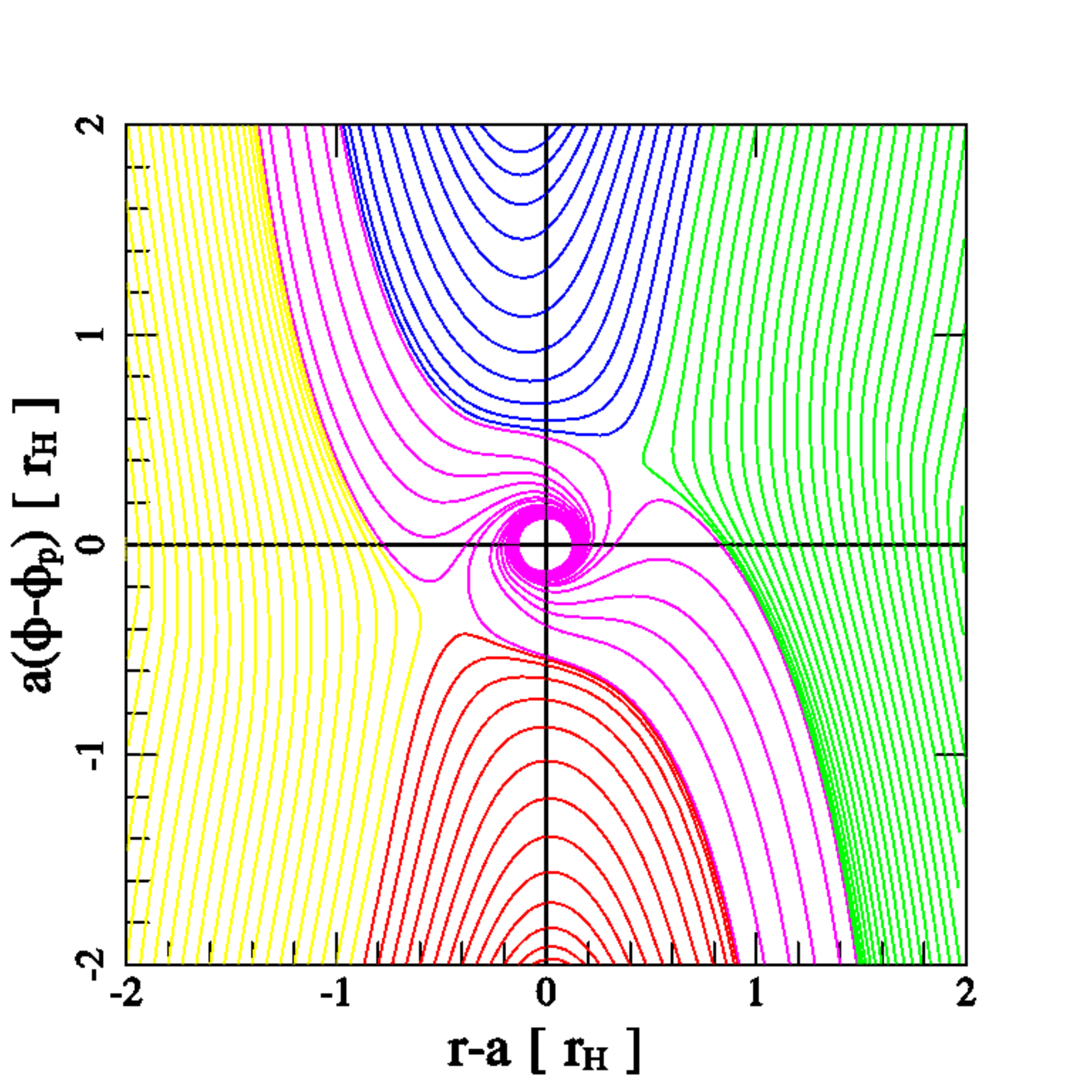}
\caption{Streamlines in the disk midplane. Compare with \figref{fig:quad} for differences between 2D and 3D flow. Yellow, red, green, and blue streamlines are assigned in the same manner as \figref{fig:quad}. Unlike \figref{fig:quad}, magenta lines are outflows away from the planet, pulled down from initially higher altitudes. They reach as close as $1.5\rs$ from the planet and are unbound.}
\label{fig:mid}
\end{figure}

\begin{figure}[]
\includegraphics[width=0.99\columnwidth]{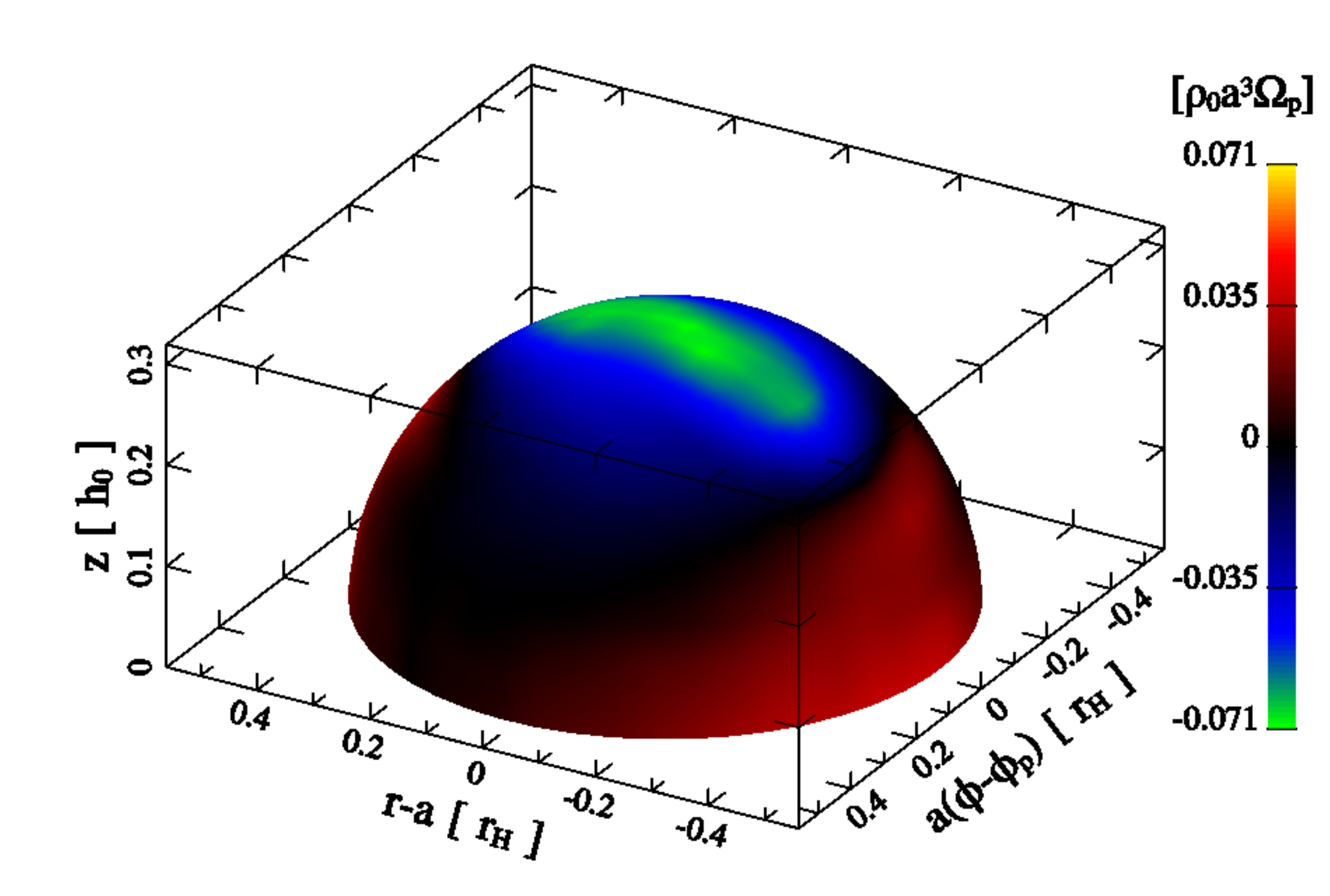}
\caption{Mass flux across the surface of a sphere centered on the planet. The sphere has a radius of $0.5\rB$. Blue and green indicate influx; red and yellow are outflux. The speed of the downward flow is about $0.7\cs$ in this plot, while the two radial outward flows in the midplane (one not visible from this viewing angle) each has a speed of $\sim 0.2\cs$, as is explained in Appendix \ref{append}. Match this figure with \figref{fig:mid} for a more complete view of the flow topology near the midplane.}
\label{fig:surf}
\end{figure}

In \secref{sec:cor_flow} we identify a rapid vertical motion in the horseshoe flow halfway through the horseshoe turn. This is a flow moving toward the planet from directly above (and below) it. Here we investigate how this vertical flow affects the planet's atmosphere. If one assumes the planet's atmosphere has an isothermal hydrostatic structure, it satisfies:
\begin{equation}
\frac{{\rm d}\eta}{{\rm d}z} = -\frac{{\rm d}\Phi}{{\rm d}z} \, ,
\label{eqn:balance}
\end{equation}
where $\eta$ is the enthalpy of the fluid, defined by ${\rm d} \eta = {\rm d} p / \rho$. This gives a vertical density profile at the planet's location:
\begin{equation}
\rho_{\rm static}(z) = \rho_0 ~ {\rm exp}\left(-\frac{z^2}{2h_0^2}+\frac{\rB}{\sqrt{z^2+\rs^2}}\right) \, .
\label{eqn:static}
\end{equation}
\figref{fig:atm} plots $\rho_{\rm static}$ together with the density profile we find from simulation. We find that near the planet, there is a large discrepancy between the hydrostatic solution and our simulation result. Within $\rB$ of the planet (the Bondi sphere), the density can be an order of magnitude less than $\rho_{\rm static}$. This is because the gas is not at rest. \figref{fig:mid} shows the streamlines in the midplane\footnote{The midplane is the bottom boundary of our simulation grid, so these streamlines approximate the midplane by following the velocity field of the bottommost cells, which are centered on $z\approx 0.0005a$, with the $z$-component velocity set to zero.} of the disk, with a color code same as \figref{fig:quad}, except for the magenta lines (see caption). Comparing to the 2D streamlines in \figref{fig:quad}, the magenta lines, which represents the flow of material from within the Bondi sphere, have a qualitatively different behavior. In 2D, the atmosphere has closed stream lines bound to the planet; in 3D, there is no static atmosphere. Rather, there is a mass inflow near the planet's poles, and a comparable outflow in the equator (\figref{fig:surf}).

So, similar to the conclusion reached by OSK15 in their study for small planets ($q_{\rm th}=0.01$), we find that the flow within the planet's Bondi sphere is not static, but instead circulates with the disk. Moreover, it is clear from \figref{fig:hs} that the flow in and out of the planet's Bondi sphere is a part of the horseshoe flow, and that the two outflowing streams in \figref{fig:mid} are simply the continuation of the inner and outer horseshoe turns that has been pulled down from above the midplane. In fact, every magenta line in \figref{fig:mid} can be traced back to a horseshoe orbit that originated from an altitude of about $0.5\sim 1 h_0$. We call this the transient horseshoe flow, which we will discuss in depth in \secref{sec:tur_flow}. Then, where is the planet's atmosphere?

There is one region where we do find 3D streamlines that do not leave the vicinity of the planet, which is a small sphere within $1.5\rs$, or $\sim0.15\rB$, of the planet. Recall that $\rs$ is equivalent to the planet's physical size. This means we are finding that the planet's atmosphere is not much larger than the pre-defined planet radius. However, it should be noted that our simulation grid resolves this region by merely 3 cells, so the flow there is not numerically accurate, creating substantial numerical viscosity that lowers the gas density. The explicit viscous force in this region also becomes large as it scales as $\rs^{-2}$, adding to the already substantial numerical viscosity. In reality, viscosity on this length scale should be much weaker than the disk viscosity, since the typical disk eddy size is $\sim h_0$, much larger than $\rs$. With a more accurate and realistic treatment, the amount of bound gas may be larger than what we measure. Other than this region, essentially all of the gas within planet's Bondi sphere are part of a more elaborate horseshoe flow.

Our results share similarities with \citet{Tanigawa2012}, who performed local, isothermal, inviscid simulations with a massive planet that has $\rH = h_0$ ($q_{\rm th}=3$), and \citet{Ayliffe2012}, who performed smoothed particle hydrodynamics simulations with 15-33 $M_{\oplus}$ planets, taking into account self-gravity and radiation. Figure 5 of \citet{Tanigawa2012} and Figure 12 of \citet{Ayliffe2012}, each showing the mass flux across a sphere of radius $0.3\rH$ around the planet, can be compared to our \figref{fig:atm}. \citet{Ayliffe2012}, in their ``post-collapse'' regime after mass accretion has slowed down, find that influx is concentrated in the vertical direction, while outflux is only in the midplane. This is in agreement with our findings. For the more massive planet that \citet{Tanigawa2012} simulated, they find both the influx and outflux of mass are more concentrated along the midplane, even though their influx is still noticeably offset to a higher latitude. This is consistent with our expectation that vertical motion plays a lesser role in the horseshoe flow for planets with $\rH >h_0$. Additionally, \citet{Tanigawa2012} reported that, similar to OSK15's results and our simulation, the gas is unbound at distances larger than $\sim0.1\rH$ from the planet. This may be indicating that the unbound nature of the gas in the Hill (for larger planets with $\rB>\rH$) or Bondi (for smaller ones) sphere is irrespective of planet mass.

OSK15 introduced the concept of replenishment timescale, $t_{\rm replenish}$, which measures the total amount of mass within the Bondi sphere, $M_{\rm BS}$, divided by the influx of mass into it, $\dot{M}_{\rm in}$. In our simulation, we have $t_{\rm replenish}\sim \Omega_{\rm p}^{-1}$. Since we do not include any sink cells to treat planet accretion, we expect the net mass flux across the Bondi sphere to be zero in steady state. We find that $\dot{M}_{\rm in}$ is balanced by $\dot{M}_{\rm out}$, the outflux of mass, to within $1\%$, with $\dot{M}_{\rm in}$ being the larger one. The Bondi sphere is therefore still slowly accumulating mass after $100 P_{\rm p}$, but on a timescale much longer than $t_{\rm replenish}$, so it is effectively in a steady state. If we lower resolution by factors of $1.2$ to $\rH/\Delta x_{\rm min}=15$ and $12.5$, the flow pattern we see in Figures \ref{fig:mid} and \ref{fig:surf} remains qualitatively the same, but $M_{\rm BS}$ decreases by $2\%$ and $5\%$, while $\dot{M}_{\rm in}$ increases by $10\%$ and $30\%$, respectively. This indicates that, not surprisingly, pressure support in the Bondi sphere is better established at a higher resolution, and our result has converged to within a level of $\sim 10\%$.

Our measurement of $t_{\rm replenish}$ is much smaller than the $t_{\rm replenish}\sim 100 \Omega_{\rm p}^{-1}$ reported by OSK15 for their $q_{\rm th}=0.01$ planet. The main difference between our result and theirs, is that the vertical motion near the planet is much faster in our case, which leads to more kinetic support, and lower density near the planet comparing to $\rho_{\rm static}$. Overall this gives a higher $\dot{M}_{\rm in}$, smaller $M_{\rm BS}$, and shorter $t_{\rm replenish}$. For comparison, OSK15 found a density profile that nearly exactly follows the hydrostatic solution (see their Figure 4), which indicates a much slower motion within the Bondi sphere, while \citet{DAngelo2003} reported supersonic vertical motion above the planet for planet masses $\gtrsim 20 M_{\oplus}$. This suggests to us that the Bondi sphere is increasingly more kinetic supported as planet mass increases.

\subsection{Transient Horseshoe Flow and Wake Vortices}
\label{sec:tur_flow}

By now, it is evident that in 3D, there exists a significant asymmetry in the flow pattern around the planet before and after the horseshoe turn. There is a new flow, which we call the transient horseshoe flow, where fluid at high altitude is pulled down toward the planet, enters its Bondi sphere, and exits radially in the midplane. These are shown by the magenta lines in \figref{fig:mid} for the midplane.

The word ``transient'' refers to the fact that this flow only performs the horseshoe turn once, even in steady state. Because of the fall in altitude, gravitational potential energy is converted to kinetic, and the flow gains radial speed as it completes its horseshoe turn. The transient flow is the portion of the horseshoe flow that has gathered enough speed to over-shoot radially and exit the horseshoe region. We analyze the radial outflow in Appendix \ref{append} and show that, due to the conservation of Bernoulli's constant, the radial outflow at $|r-a|=\rH$ has a speed of $|u_{\rm r}|\approx 0.6\cs$, while we measure  $0.2\cs$ to $0.4\cs$ in our simulation. Appendix \ref{append} also demonstrates that this outflow speed is independent of planet mass in the limit $q_{\rm th}\gg1$, but is slower for lower mass planets.

\begin{figure*}[]
\includegraphics[width=1.99\columnwidth]{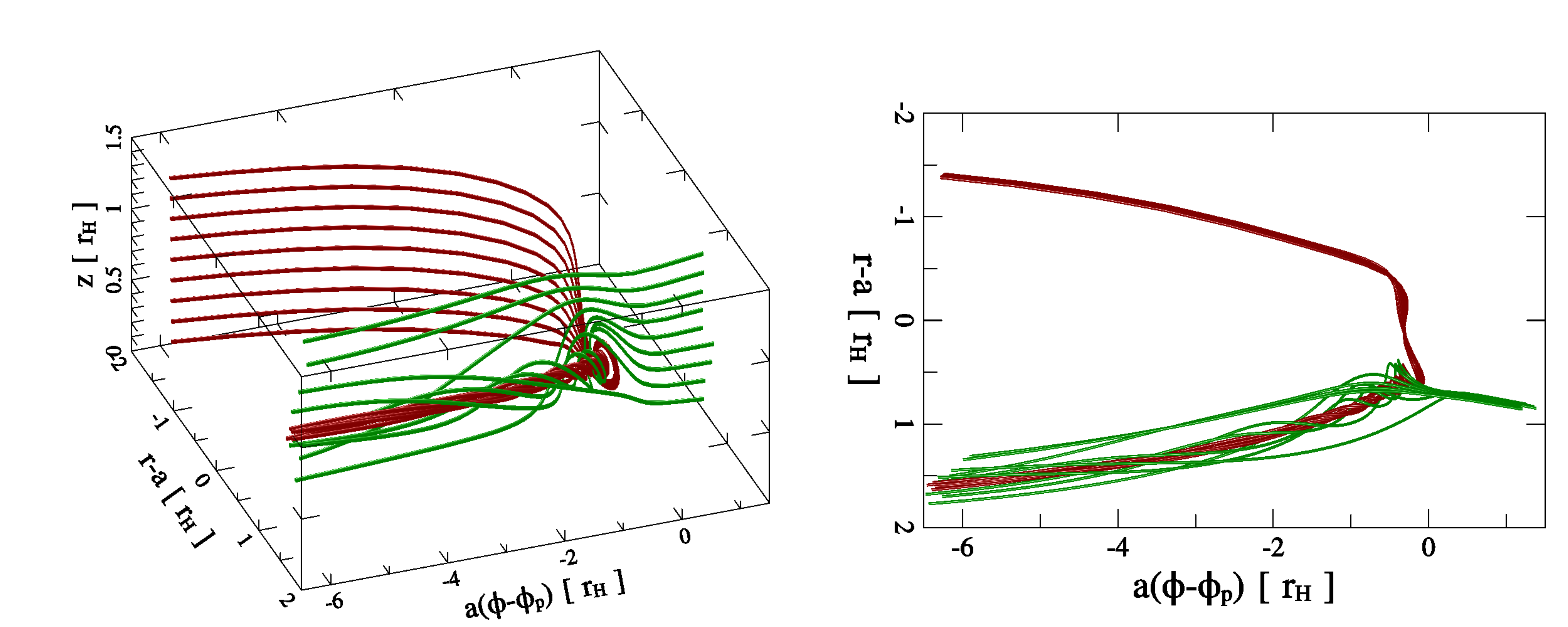}
\caption{Streamlines at the boundary of the horseshoe region. The red lines are inner horseshoe flow; green are outer disk flow. After a close approach to the planet, the red streamlines turn around and descend to the midplane of the disk, sliding underneath the green streamlines. Green lines in higher altitude simply enters the horseshoe region, while lower ones are mixed with the red lines. Similarly, but not shown here, this also happens between the inner horseshoe flow and the outer disk flow.}
\label{fig:mix}
\end{figure*}

Since material in the horseshoe region is being lost to the transient flow, it must be replenished. This comes from the disk flow lying outside of the horseshoe region: as the transient flow completes its horseshoe turn, it becomes vertically compressed, creating a low pressure region above it, which attracts the high altitude flow right next to the horseshoe region to move in (\figref{fig:mix}). This establishes an exchange of material between the horseshoe region and the disk , as the red and green streamlines in the figure wrap around each other. 

\begin{figure}[]
\includegraphics[width=0.99\columnwidth]{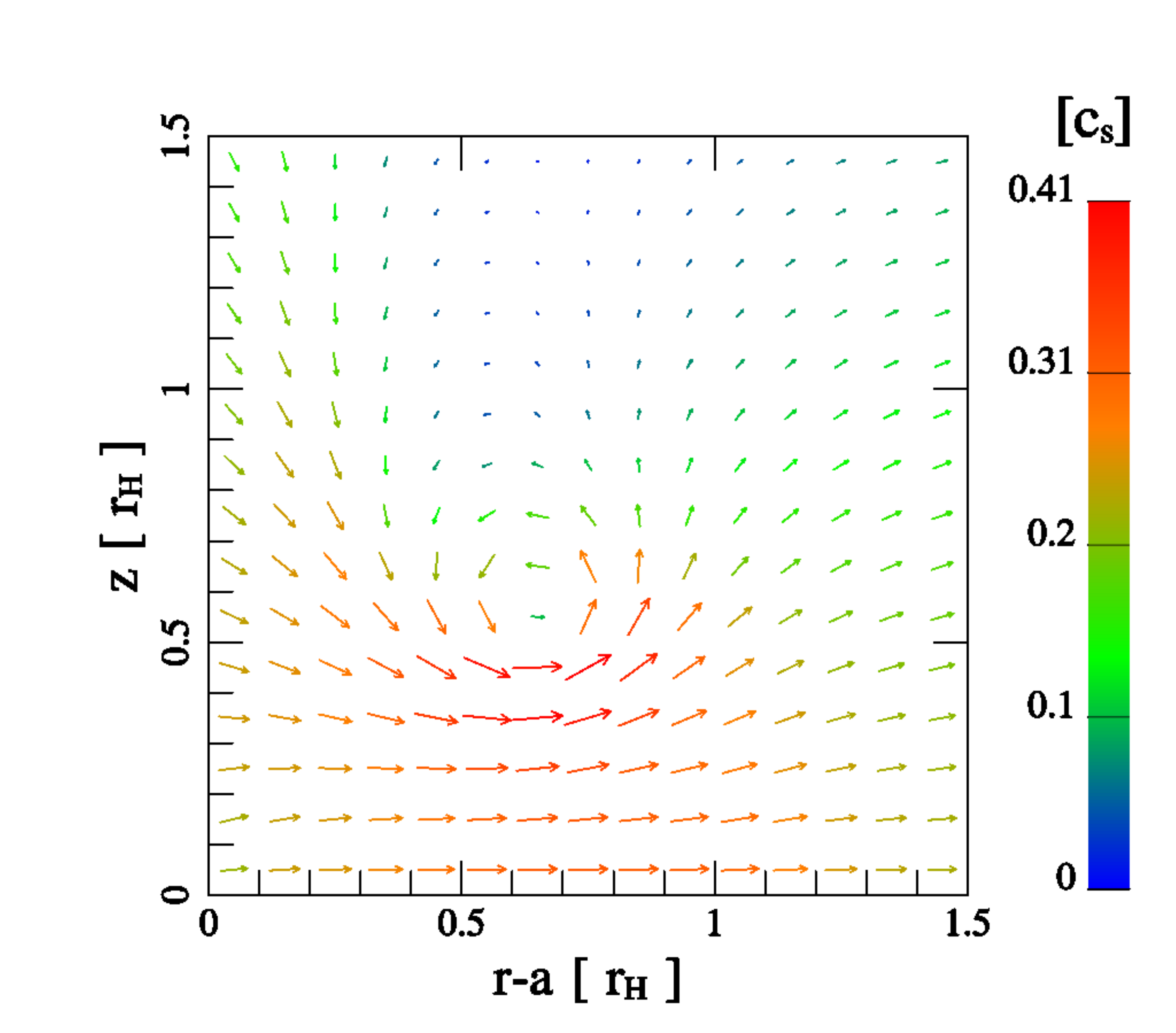}
\caption{Velocity field on a meridional plane at $\phi=\phi_{\rm p}-0.5\rH/a$. The color of the arrows indicates the speed. The fastest radial flow speed is $\sim0.4\cs$ in this plot. The vortex roll-up occurs between $0.5\sim1\rH$ away from the planet, and about $0.5\rH$ above the midplane. The size of the vortex core is about $0.1\sim0.2\rH$.}
\label{fig:rz}
\end{figure}

Finally, the vertically compressed transient flow needs to decompress as it settles into the disk. This is done through a meridional circulation triggered by a ``vortex roll-up'', illustrated by \figref{fig:rz}. 
The same meridional flow was also identified by \citet{Morbidelli2014} in their 3D simulation for a Jupiter-mass planet. As the transient flow over-shoots the horseshoe region, it gets deflected upward by the midplane disk material, and rolls over to decompress itself, causing the vortex roll-up. This phenomenon is similar to the behavior of the heads of plumes of fluid intruding into stationary fluid, which tend to roll at the edges to form mushroom-head shapes. However, the closest analogy is the formation of wingtip vortices in aerodynamics of finite-span wings, where higher pressure air underneath most of the wing length begins to move down (in a downwash), but near the wingtip also moves sideways and circulates around the tip to move into the low pressure region above the wing. The downwash and the associated aerodynamical lift are due to the circulation induced around the wing by vortex lines attached to the airfoil spanwise (Joukovsky theorem 
states the proportionality of lift and circulation). By Kelvin's circulation theorem, the vorticity $\omega$, a solenoidal, divergence-free field, cannot simply end at the ends of the finite-span airfoil. Instead, the vortex line remains continuous and preserves the circulation; it only changes direction and is shed from the wingtips into the wake as two free wingtip vortices separated roughly by one wingspan.   

In the protoplanetary disk, we have a close analogy: the gas in the transient horseshoe flow after a passage near the planet is torqued by its gravity and forms two outflow plumes, one directed toward the star 
and one in the opposite direction. The vertical extent of this flow is $\sim\rH$, which plays the role of the aerodynamical wingspan of airfoil; two equivalent airfoils of this length would be positioned vertically and separated radially, also by a distance on the order of a few $\rH$. Therefore, the planet's torque on the gas necessarily creates radial plume-like flows of gas, but also a total of four, alternately turning, nearly horizontal vortex lines shed into the disk flow near the interface between the horseshoe and disk regions. \figref{fig:mix} shows one of these vortices. We call them linear wake vortices, as a reminder that compared with their core diameters, they can be very long, as they are carried by the disk flow.   

Furthermore, the equation of vortensity \eqnref{eqn:vortensity} supplemented by viscous dissipation
term, which we dropped before, can be written as 
\begin{equation}
\frac{{\rm D}\bm{\xi}}{{\rm D}t} =  \left(\bm{\xi}\cdot\bm{\nabla}\right)\bm{u} +\nu \bm{\nabla}^2\bm{\xi} \,,
\label{eqn:vortensity2}
\end{equation}
where $\bm{\xi}$ is the vortensity given by \eqnref{eqn:vortensity_def}.  
The evolution of vortensity in the core of the vortex can be deduced from this equation, after noticing that 
vortensity and velocity $\bm{u}$ are almost parallel and directed along the vortex line.
 
The first term on the RHS of the above equation is thus the gradient of longitudinal velocity 
of material along the vortex core multiplied by $\xi$.  
As shown in \figref{fig:mix}, the roll-up of the wake vortices happens near the stagnation region, where azimuthal motion with 
respect to the planet is slow. After the vortex core forms, it is carried 
at an increasing speed into the disk flow, therefore the velocity gradient is first strongly positive, 
and further from the planet drops to zero, as the flow becomes an azimuthally non-accelerated disk flow. 
While the flow accelerates, the full time derivative of vortensity grows, strengthening the
wake vortex exponentially (in the direction of vortex line, spatial derivative of velocity equals ${\rm D}\ln\bm{\xi}/{\rm D}t$). Only when the vortex starts freely coasting away from the planet, several $\rH$ behind its roll-up region, 
the second term in the equation representing viscous spreading becomes dominant,
diffusing the vortex and weakening the vortensity of its core. 

Viscous dissipation of the core has an associated timescale equal to the squared diameter $d^2$
of the core divided by $\nu$, during which fluid elements are carried in azimuth by differential rotation 
through a distance $\lambda \sim (d^2/\nu)\Omega_{\rm p} \rH$. If $d\sim 0.1\rH$ as seen in \figref{fig:rz}, $\rH = 0.017a$
appropriate for our $5M_\oplus$ planet, and $\nu = 10^{-6}a^2 \Omega_{\rm p} $, then the estimated distance 
is $\lambda \sim 3\rH$. This is an estimate of the distance on which viscosity in our simulation will 
double or e-fold the initial diameter by assumption equal to $0.1\rH$. The freely flowing line vortex can spread and 
effectively disappear after a few length scales $\lambda$. This estimate agrees very well with our numerical 
simulations: vortices are observed to weaken substantially past $10\rH$ behind the planet (\figref{fig:fvor}). 
It is intriguing to ask what will happen to this vortex if the disk is inviscid, which we will investigate in a future paper.

\section{Torque on the Planet}
\label{sec:tor}

As mentioned in \secref{sec:parameters}, our disk has a flat 2D vortensity profile, $\omega/\Sigma = $ constant, which is predicted by 2D analysis to have a vanishing corotation torque. Since vortensity is a conserved quantity along a 2D streamline, a fluid element performing a horseshoe turn will always maintain a density the same as its surrounding if the background vortensity profile is constant; hence there cannot be a density difference along any given $r$ between the inner and outer horseshoe flow, and no corotation torque can be generated. As a result, the only remaining torque is the differential Lindblad torque, which, by the 3D linear calculations by \citet{Tanaka2002}, is $-2.19T_0$, where $T_0 = \Sigma_0 a^4 \Omega_{\rm p}^2 q^2 (h_0/a)^{-2}$, for an isothermal disk with $\Sigma\propto r^{-3/2}$.

The above results may not be applicable in 3D for two main reasons. First, while the 2D vortensity $\omega/\Sigma$ can be a constant in the disk, the 3D vortensity $\omega/\rho$ depends on $z$. For instance, in our model, $\omega/\rho\propto r^{3/2}$ in the midplane even though $\omega/\Sigma$ is constant. This is because $\rho(z=0)\propto r^{-3}$ (see \eqnref{eqn:rho}). Second, and more importantly, vortensity is a conserved quantity along a streamline in 2D, but not in 3D, as is evident in \eqnref{eqn:vortensity}. In fact, \figref{fig:fvor} already shows that planar vorticity is generated after the widest horseshoe turns. This section aims to investigate how the corotation torque behaves in 3D.

\begin{figure}[]
\includegraphics[width=0.99\columnwidth]{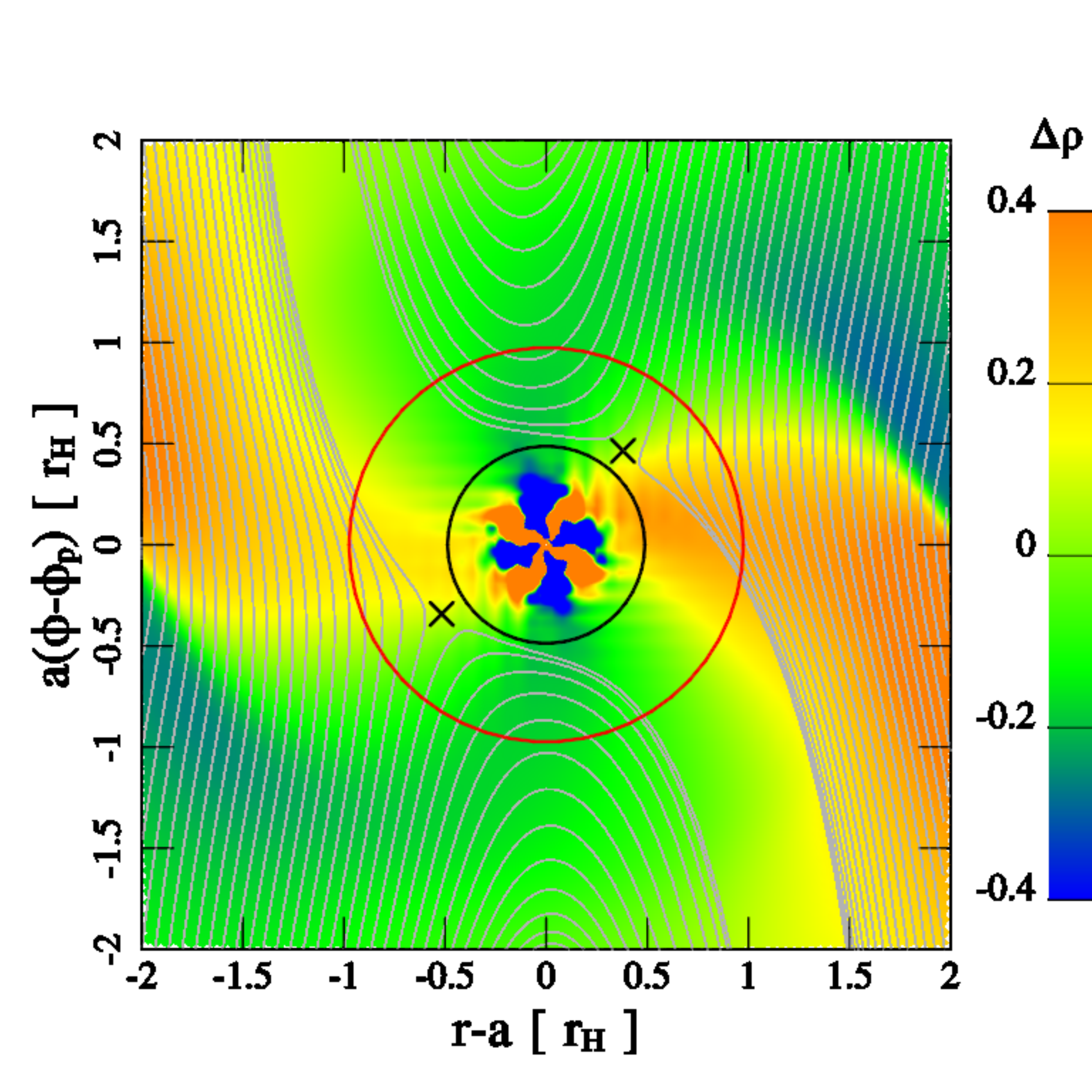}
\caption{The midplane non-axisymmetric density distribution around the planet, scaled by the background density (see \eqnref{eqn:rho_asym}). The gray lines are streamlines in \figref{fig:mid}, except with the magenta lines omitted. The crosses mark the stagnation points. They are located at $\{x,y\}=\{-0.47,-0.36\}$ and $\{0.42,0.42\}$, in units of $\rH$. This is different from the 2D case (\figref{fig:quad}), where the stagnation points lie close to $\phi_{\rm p}$. The black circle has a radius of $0.5\rB$. Because of the non-physical four-armed spiral inside the black circle, we exclude this region from our torque calculation. The red circle's radius is $\rB$, corresponding to the sphere where the red curve in \figref{fig:dis} is computed.}
\label{fig:stag}
\end{figure}

Before computing the planetary torque in our simulation, we first inspect the density structure near the planet. \figref{fig:stag} plots the midplane density scaled by the background density and with the axisymmetry density about the planet removed:
\begin{equation}
\Delta \rho = \rho \left(\frac{r}{a}\right)^3 - \frac{1}{2\pi}\int_{0}^{2\pi} \rho ~{\rm d}\phi' \, ,
\label{eqn:rho_asym}
\end{equation}
where $\phi'$ is the azimuth in a polar coordinate centering on the planet. The axisymmetry part of the density does not contribute a net torque, so it can be safely removed for clarity. One notable problem we can see in \figref{fig:stag} is the four-armed spiral around the planet. It is a numerical artifact due to the local Cartesian grid geometry around the planet. This problem was also identified by \citet{Ormel2015a}. It can be reduced by sufficiently resolving $\rs$, which unfortunately is not the case for us, since we only resolve $\rs$ by about 2 cells. The four-armed spiral introduces an artificial torque on the planet that needs to be removed. We therefore exclude the torque contribution from within $0.5\rB$ of from the planet, shown in \figref{fig:stag} as the black circle.

\begin{figure}[]
\includegraphics[width=0.99\columnwidth]{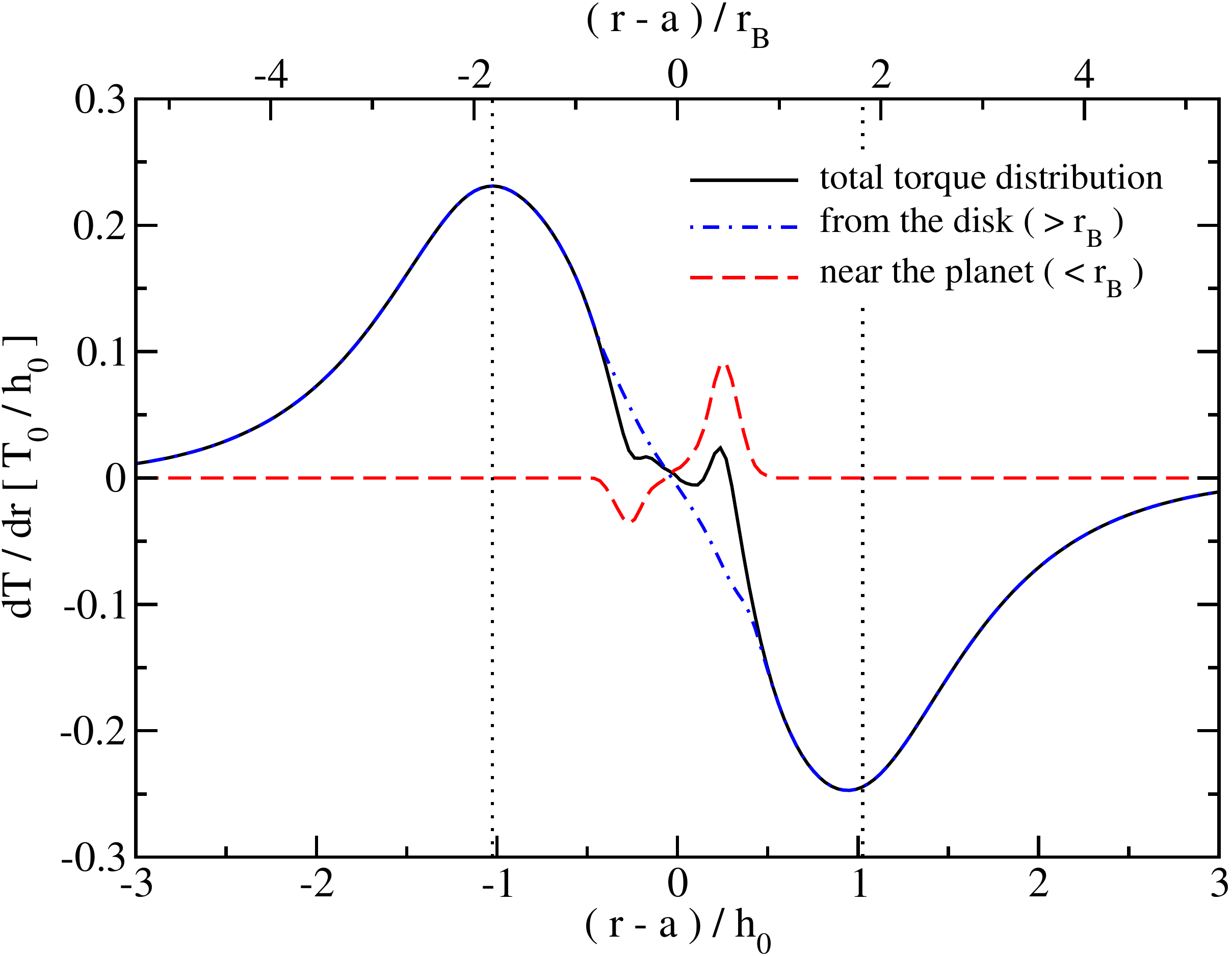}
\caption{Torque distribution as a function $r$. The black solid curve is the total torque distribution, and is equal to the sum of the red dashed and blue dash-dotted curves. The red curve only includes contribution from within a sphere of $1~\rB$ around the planet (see \figref{fig:stag}), while the blue curve includes the rest of the disk. The two black dotted lines draw the boundaries of the horseshoe region. The two blue bumps at $\pm h_0$ correspond to the outer and inner Lindblad torques; and the two red bumps near the planet are caused by the stagnation point offsets seen in \figref{fig:stag}.}
\label{fig:dis}
\end{figure}

\begin{figure}[]
\includegraphics[width=0.99\columnwidth]{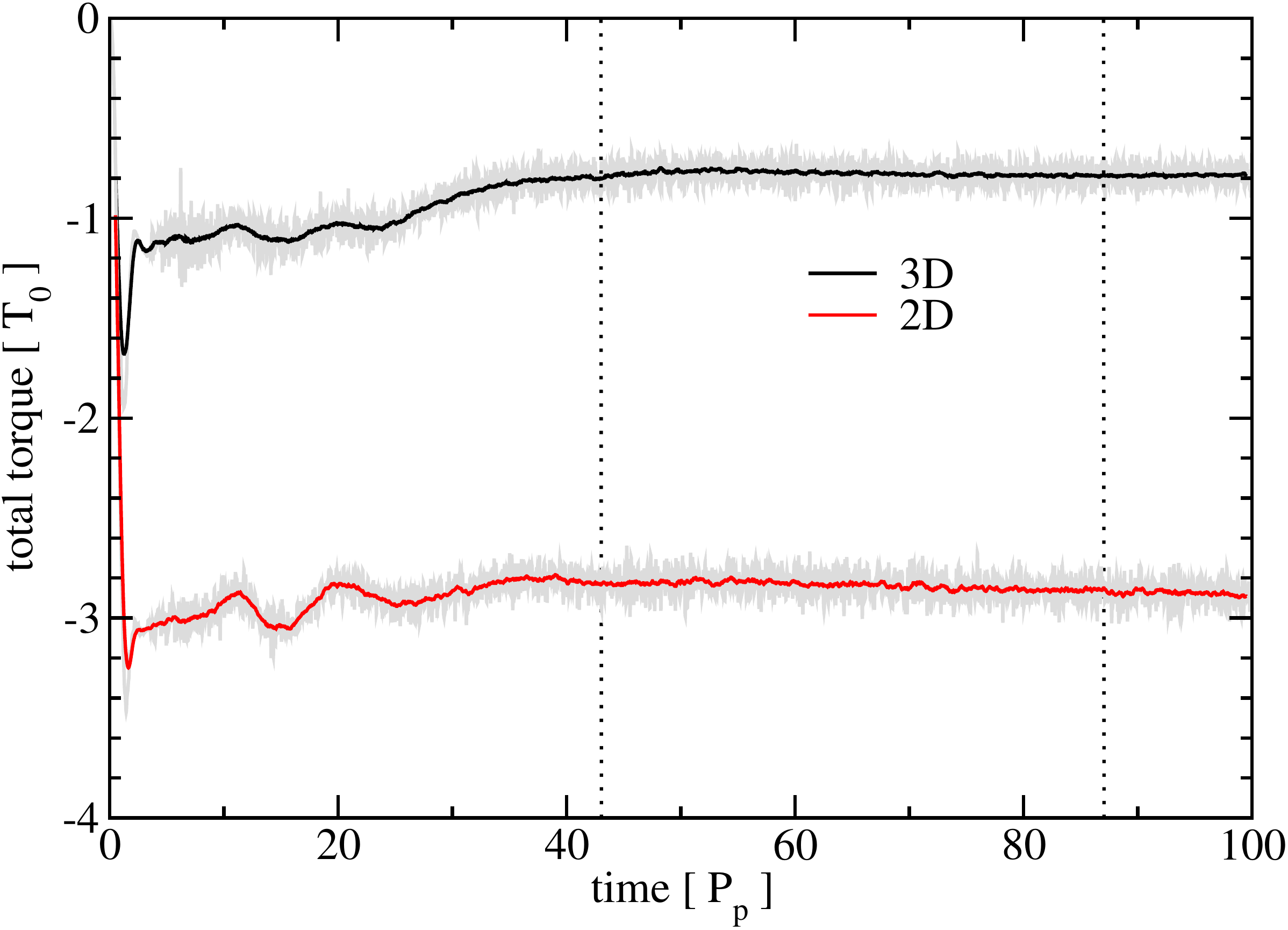}
\caption{Net torque on the planet as a function of time. The black curve is our 3D torque measurement; red is 2D. This 2D case shares the same setup as the 3D one, except for $\rs=0.3h_0$. Both curves are running-time-averages over $1P_{\rm p}$. The instantaneous values of the the torques are shown as the gray shades around each curve. The vertical dotted lines mark the libration time of the horseshoe orbit. The first dotted line is at $1~t_{\rm lib}=43~P_{\rm p}$, and second one is $2~t_{\rm lib}=86~P_{\rm p}$.}
\label{fig:time}
\end{figure}

\begin{figure}[]
\includegraphics[width=0.99\columnwidth]{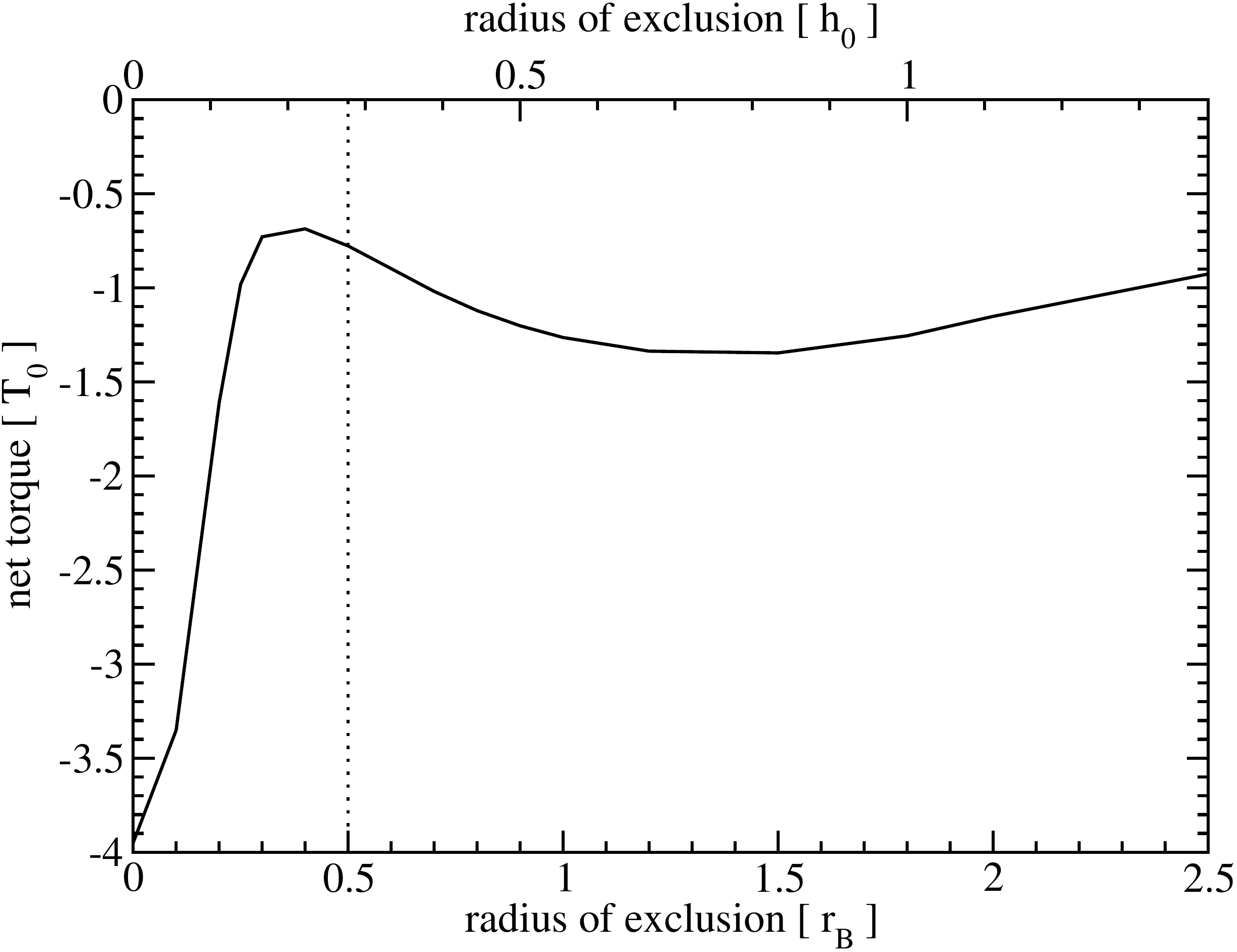}
\caption{Net torque on the planet as a function of the radius of the excluded sphere centered on the planet, shown as the black solid curve. This plot, together with \figref{fig:stag}, shows the non-physical four-armed spiral residing within $\sim0.4\rB$ from the planet contributes a significant amount of torque that should be excluded from our calculation. The black dotted line labels the radius of exclusion we use, $0.5\rB$, corresponding to the black circle in \figref{fig:stag}.}
\label{fig:exclude}
\end{figure}

Now we compute the torque. \figref{fig:dis} plots the torque distribution ${\rm d}T/{\rm d}r$, which is the amount of torque on the planet by the disk at a given $r$:
\begin{equation}
\ddr{T} = \int_0^{2\pi} \int_{-\infty}^{\infty} \rho \frac{\partial\Phi}{\partial\phi} ~ {\rm d}z ~ {\rm d}\phi \, ,
\label{eqn:tor}
\end{equation}
where $T$ is the net torque on the planet. \figref{fig:time} plots the net torque as a function of time, demonstrating that our measurements have converged with time. \figref{fig:exclude} shows how the net torque depends on our choice of the excised region's radius. We find the radius needs to be at least $0.4\rB$, or about 7 grid cells, to fully remove contribution from the non-physical four-armed spiral, and our choice of $0.5\rB$ safely accomplishes that without excessively removing contribution from the disk. We further divide the torque into two components: one from within the planet's Bondi sphere (red curve; contribution from $|\bm{r}-\bm{r}_{\rm p}|<\rB$, represented by the red circle in \figref{fig:stag}), and one from the rest of the disk (blue curve). While the blue curve has the characteristic shape of the Lindblad torque distribution, the red curve is not a well-known feature. If we integrate each curve, the red curve gives a torque of $+0.50T_0$, while the blue one gives $-1.27T_0$. The net torque is therefore $-0.77T_0$. This is significantly weaker than the result from linear calculation ($-2.19T_0$).

We also perform a 2D simulation to show this result is unique in 3D. The 2D setup is identical to 3D, except we set $\rs=0.3h_0$, since in \secref{sec:cor_flow} we find this smoothing length produces the best matching horseshoe width. Our 2D torque is $-2.86T_0$ (\figref{fig:time}), comparable to the 3D value from linear calculation. For comparison, \citet{DAngelo2003} also found 3D torques are about one order of magnitude weaker than 2D when the planet mass is around 10 $M_{\oplus}$.

Similar to how the blue curve has two bumps near $r-a=\pm h_0$, and red curve also has two separate bumps in the inner and outer disk, but with reversed signs compared to the blue. We will refer to this behavior as ``torque reversal''. This was also seen by \citet{DAngelo2010} in their Figure 15 where they simulated planets with masses larger than a few $M_\oplus$. This fact, together with \figref{fig:stag}, provides clues to the nature of this torque. In \figref{fig:stag} we marked two stagnation points, where $|\bm{u}|=0$. We see that in the outer disk, the stagnation point is slightly above $\phi_{\rm p}$, while it is below $\phi_{\rm p}$ in the inner disk. This is significantly different from the 2D flow pattern (see \figref{fig:quad}), where both stagnation points lie much closer to $\phi_{\rm p}$. Because of this offset, high density regions are created at $\phi$ larger (less) than $\phi{\rm_p}$ in the outer (inner) disk, thus generating a positive (negative) torque near the planet. However, it remains unclear to us why the net contribution from the red curve is positive, i.e., the offset in the outer disk is larger than the inner disk, conveniently reducing the net migration rate. 

We also note that even if we ignore the red curve, the blue curve still only contributes a torque of $\sim-1.3T_0$, significantly weaker than $\sim -2.2T_0$ from either linear calculations or 2D simulations. This may have to do with the fact that the blue curve contains both Lindblad and corotation torque. From \figref{fig:dis} we can see that a large fraction of the torque distribution coincides with the horseshoe region. This prevents us from distinguishing which part belongs to the corotation torque, and which part is the Lindblad torque. However, we can measure the corotation torque with a different method, separate from \figref{fig:dis}. 

\begin{figure*}[]
\includegraphics[width=1.99\columnwidth]{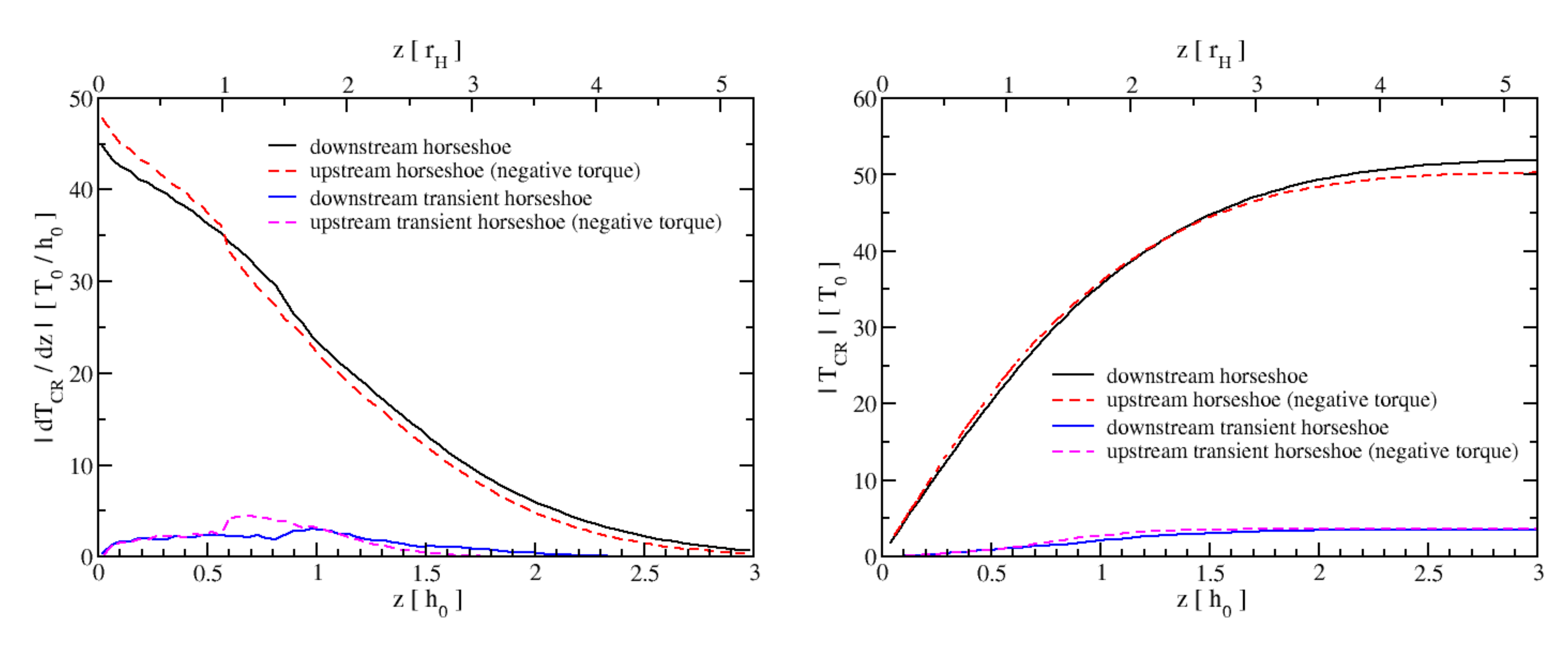}
\caption{Magnitude of the differential corotation torque on the left panel, and magnitude of the cumulative corotation torque on the right, both as functions of height above the midplane. Black sold curve represents contribution from the outer horseshoe flow; red dashed curve corresponds to the inner one. Similarly, blue solid and magenta dashed curves are contributions from the outer and inner flows respectively, but are transient flows that exit the horseshoe region after one turn. Contributions from inner flows are negative in value. On the left panel, one can see that while the regular horseshoe flow provides the strongest torque near the midplane, the transient flow comes from an altitude of $\sim h_0$. On the right panel, it shows that overall the outer horseshoe flow generates a larger torque than inner. The sum of all 4 components is $1.5T_0$.}
\label{fig:CR}
\end{figure*}

We follow a fluid element's motion starting at $\phi=\phi_{\rm p}-1$ for the inner flow, or $\phi_{\rm p}+1$ for the outer, until it completes its horseshoe turn and returns to its starting azimuth. $\Delta l$ is then the difference in the fluid's specific angular momentum between its start and end points. Combining this with the flow rate in the horseshoe region, the corotation torque, $T_{\rm CR}$, is:
\begin{align}
T_{\rm CR}(z) &= T_{\rm CR,i}(z) + T_{\rm CR,o}(z) \\
 &= \left.\int_{a-w_{\rm i}}^a \int_{-z}^{z} \rho |u_{\phi}| \Delta l ~ {\rm d}z' ~ {\rm d}r\right|_{\phi=\phi_{\rm p}-1} \\
&+ \left. \int_a^{a+w_{\rm o}} \int_{-z}^{z} \rho |u_{\phi}| \Delta l ~ {\rm d}z' ~ {\rm d}r \right|_{\phi=\phi_{\rm p}+1} \, ,
\label{eqn:cor}
\end{align}
where $T_{\rm CR,i}$ is the corotation torque due to the inner horseshoe flow, $T_{\rm CR,o}$ is the outer one; $w_{\rm i}$ and $w_{\rm o}$ are the horseshoe half-widths for the inner and outer flow respectively. Furthermore, we can separate the contribution from the transient horseshoe flow by identifying streamlines that settle outside of the horseshoe region. \figref{fig:CR} plots the differential torque $|{\rm d}T_{\rm CR}/{\rm d}z|$ on the left panel and cumulative torque $|T_{\rm CR}|$ on the right. Here $z$ refers to starting position of the streamline, not where the torque exchange happens. A caveat with this method is that our velocity field is time-averaged over just $1 P_{\rm p}$, whereas the libration time for these horseshoe orbits are much longer. Nonetheless, because \figref{fig:time} shows the $1~P_{\rm p}$ time-averaged net torque has little fluctuation over a libration time, this method should be sufficiently accurate for our purpose.

We find that the one-sided torques, $T_{\rm CR,i}$ and $T_{\rm CR,o}$, each has a magnitude of $\sim 50T_0$. The transient horseshoe contributes $\sim6\%$ of that torque. $T_{\rm CR,o}$ is stronger than $T_{\rm CR,i}$ by $\sim3\%$, and the net corotation torque is $T_{\rm CR}\sim1.5T_0$. This sufficiently accounts for the difference between our measured net torque ($-0.77T_0$) and the expected differential Lindblad torque ($-2.19T_0$), which strongly suggests that the corotation torque is responsible for both the torque from the red curve and the reduction of disk torque in the blue curve. Recalling that our disk profile would have zero net corotation torque in 2D, our result accentuates the difference between 3D and 2D torques. The magnitude and sign of the net corotation torque is closely related to the asymmetry in the stagnation point offsets, and will be the topic for a future paper.

\citet{DAngelo2010} calculated the fully nonlinear 3D torque on a planet with $q\sim 3\times 10^{-6}$ and a disk profile similar to ours, and found the net torque on the planet to be $T = -2.29T_0$, consistent with previous 2D results, implying the corotation torque is negligible when $\Sigma\propto r^{-3/2}$. Their simulations differ from ours in two ways. First, our planet is 5 times more massive, and over 20 times larger in terms of $q_{\rm th}$. Therefore one can expect our result to deviate somewhat from linear calculations. Second, their simulation is in the regime where $t_{\nu}<t_{\rm lib}$, while ours has $t_{\nu}>t_{\rm lib}$ (see \secref{sec:parameters}). We believe the second point is the main cause for the discrepancy between their result and ours. In the next section, we will simulate a more viscous disk, and check whether we can recover their result.

\section{Dependence on Viscosity}
\label{sec:viscosity}

\begin{figure*}[]
\includegraphics[width=1.99\columnwidth]{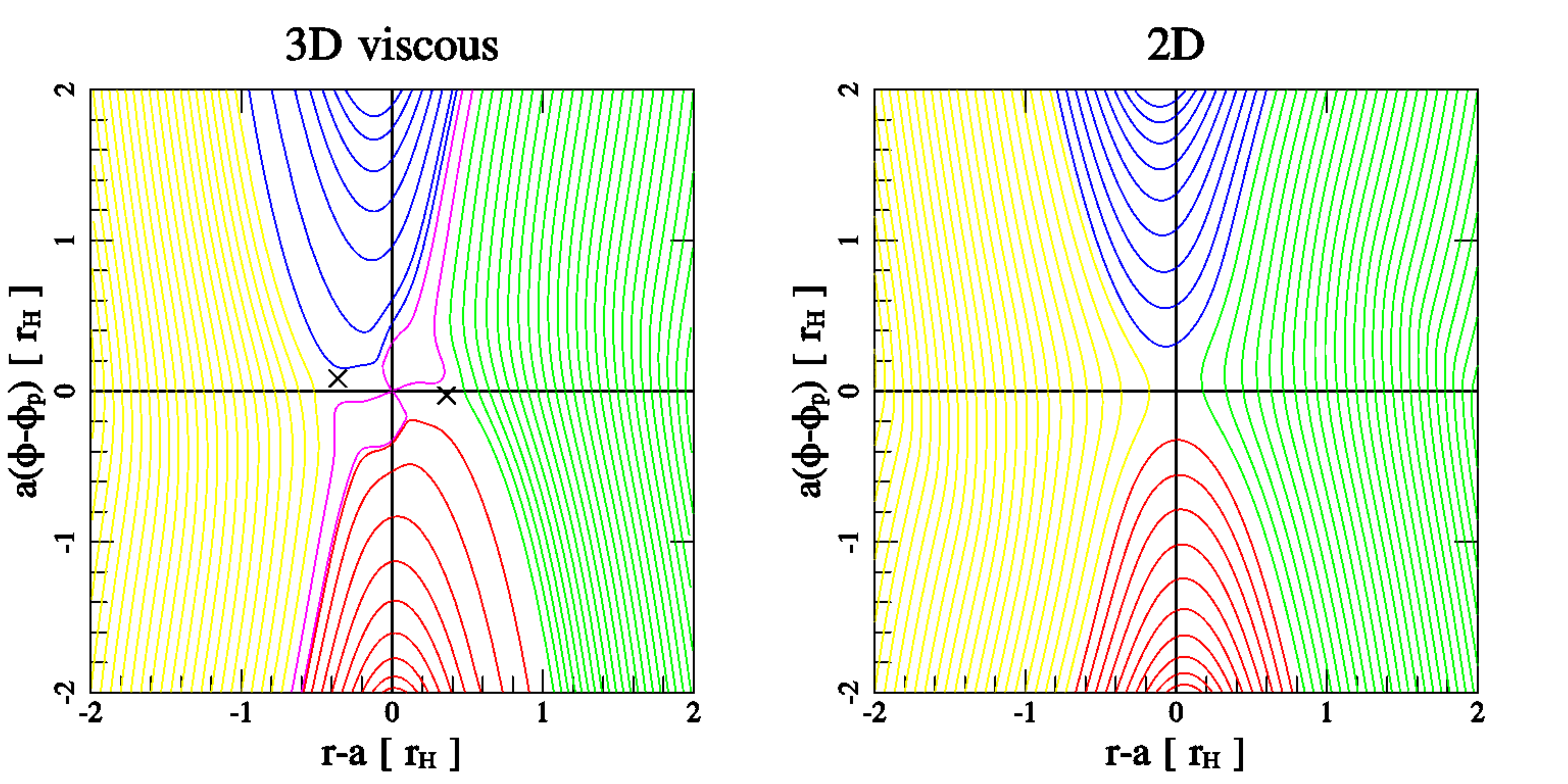}
\caption{Midplane streamlines for the 3D viscous case on the left panel, and a 2D case on the right. This 2D case here is the same as the one in \figref{fig:time}. Comparing to \figref{fig:mid}, the flow topology in the 3D viscous case is less asymmetric about $r=a$, and therefore is more similar to the 2D case on the right. The stagnation points, labeled as crosses on left, are located at $\{x,y\}=\{-0.36,0.08\}$ and $\{0.36,0.03\}$, in units of $\rH$. In the 2D case, the large smoothing length ($\rs=0.3h_0$) results in the loss of both stagnation points. Like \figref{fig:quad}, there is a stagnation point at the planet's location on both the left and right panels, which we omit to label.}
\label{fig:stream}
\end{figure*}

\begin{figure}[]
\includegraphics[width=0.99\columnwidth]{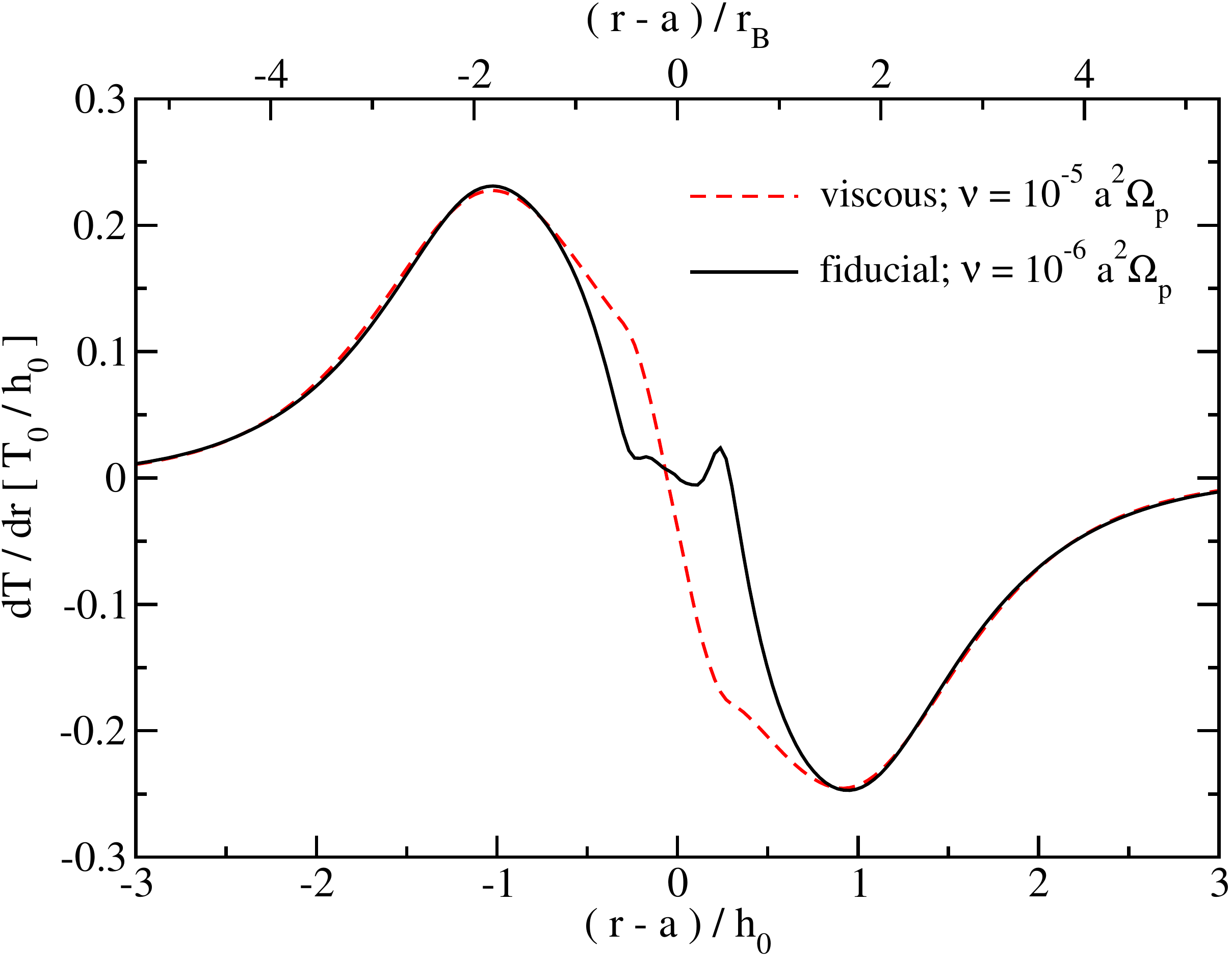}
\caption{Torque distribution as a function $r$. The black solid curve is identical to the black curve in \figref{fig:dis}. The red dashed curve is the torque distribution of our viscous case. Note the torque reversal near the planet does not exist for the red curve. This is consistent with \figref{fig:stream}, where we see the stagnation points no longer have a large azimuthal offset.}
\label{fig:vis}
\end{figure}

We increase the viscosity in our model to $\nu=10^{-5}a^2\Omega_{\rm p}$ ($\alpha\sim0.01$) and investigate how this will affect the flow pattern and the torque on the planet. We will refer to this case as the ``viscous'' case. Under this setup, the viscous diffusion timescale across the horseshoe region $t_{\nu}$ is $\sim15P_{\rm p}$, shorter than $t_{\rm lib}\sim43 P_{\rm p}$. As a result, most of the gas in the horseshoe region can only complete one horseshoe turn before being removed due to the background viscous flow. This prevents the flow pattern described in \secref{sec:results} from fully setting up. We find that in this scenario, the flow has significantly reduced vertical variation, and becomes more similar to the 2D case. 

\figref{fig:stream} plots the midplane streamlines similar to \figref{fig:quad} and \ref{fig:mid}, but for the viscous case and a 2D simulation with $\rs=0.3h_0$ (unlike \figref{fig:quad}, which has $\rs=0$). In contrast to \figref{fig:mid}, the magenta lines in the viscous case are now inflowing streams entering the Bondi sphere that converges at the stagnation point where the planet is located. They correspond to the gas attempting to stably orbit the planet, but loses angular momentum (with respect to the planet) too rapidly due to both numerical and explicit viscosity. The speed of these magenta lines is very slow close to the planet, less than $1\%$ of $\cs$, suggesting much of the flow diverges from the midplane before ever reaching the planet. As a result, most of the influx of mass is still from the vertical direction above the planet. On the other hand, the asymmetry about $r=a$ is much reduced in the viscous case. As we have shown in \secref{sec:tur_flow}, the asymmetry is caused by the vertical flow near the planet, so, not surprisingly, we find significantly reduced vertical motion. The speed at $0.5\rB$ above the planet is $0.1\cs$, 7 times slower than our fiducial case (\secref{sec:ver_flow}). 

The more symmetric streamlines also mean the stagnation points now lie much closer to $\phi_{\rm p}$, seen on the left panel of \figref{fig:stream}. This reduces the net torque from the horseshoe region. \figref{fig:vis} plots the torque distribution of the viscous case as well as that from \figref{fig:dis} for comparison. It is evident that the torque reversal seen before between the blue and red curve in \figref{fig:dis} has now been erased by the extra viscosity. The net torque on this planet is $-2.2T_0$, while the 2D case on the right panel of \figref{fig:stream} has a torque of $-2.86T_0$. These are in agreement with linear calculations of the differential Lindblad torque in 3D \citep{Tanaka2002} and 2D \citep{PaardekooperP2008}, respectively. We therefore conclude that, when $t_{\nu}<t_{\rm lib}$, the 3D flow field will become similar to 2D, and yield a similar torque on the planet as well. 

\citet{DAngelo2013} performed 3D simulations of disk-planet interaction including radiative transfer and realistic opacity. They implemented a viscosity of $\nu=4\times10^{-6}a^2\Omega_{\rm p}$, and an initial scale height\footnote{There is some ambiguity in the definition of a scale height in their work, because of the non-trivial temperature profile in their radiation-hydrodynamics treatment. See their Section 4 for details.} of $h_0\sim 0.06a$ at the planet's location. If we assume $w\sim1.2~a\sqrt{aq/h_0}$, then their model for a $5M_{\oplus}$ planet is expected to have $t_{\nu}\sim14P_{\rm p}$, much shorter than $t_{\rm lib}\sim70P_{\rm p}$. This places their model comparable to our viscous case. Comparing the left panel of our \figref{fig:stream} to their Figure 10, one can see that the flow patterns are largely similar, and neither of them show significant radial outflow at the midplane from within the Bondi sphere. This suggests that, in comparison to our fiducial case, the vertical inflow does not penetrate as deep into the Bondi sphere, which is consistent with the fact that the inflow speed is also much slower.

Finally, we note that $t_{\nu}>t_{\rm lib}$ is not only a criterion for disk viscosity, but also sets a lower limit for the planet mass. Namely, it can be rewritten as:
\begin{equation}
q > \left(\frac{8\pi}{3}\right)^{\frac{2}{3}} \left(\frac{h_0}{a}\right) \left(\frac{\nu}{a^2 \Omega_{\rm p}}\right)^{\frac{2}{3}} \, ,
\end{equation}
if we approximate $w\approx a\sqrt{aq/h_0}$. Together with the fact that 3D effects are most relevant for embedded planets: $\rH<h_0$, which can be rewritten as:
\begin{equation}
q < 3\left(\frac{h_0}{a}\right)^3 \, ,
\end{equation}
these two criteria bracket the range of planetary mass where we expect the 3D flow field to deviate most from 2D.

\section{Discussion and Conclusion}
\label{sec:discussion}
We present a detailed picture of the 3D flow topology near an embedded planet on a fixed circular orbit, extracted from 3D hydrodynamical simulations of a $\sim 5 M_{\oplus}$ planet interacting with a circumstellar disk. Our simulations are run with our GPU hydrodynamical code \texttt{PEnGUIn}, on a single desktop computer equipped with 3 GTX-Titan graphics cards. We found that the 3D modifications to the horseshoe flow have a significant influence on both the density structure in the planet's Bondi sphere, and the torque exerted on the planet. Below we give a summary of the 3D horseshoe flow:

\begin{itemize}
\item[(1)] At the onset of a horseshoe turn, before the close encounter with the planet, the flow is columnar. This results in a nearly constant horseshoe half-width $w$ in $z$ (\figref{fig:xs}).
\item[(2)] While a fraction of the horseshoe flow continues in columnar form after the turn, the widest portion is pulled toward the midplane and fall directly on top of the planet (\figref{fig:hs}). This flow plummets deep into the planet's Bondi sphere (\figref{fig:surf}).
\item[(3)] The release of potential energy from the fall results in the flow exiting the Bondi sphere near the midplane at a speed of order $\cs$ (Appendix \ref{append}). Symmetry of the horseshoe streamline about $r=a$ is broken (\figref{fig:mid}). Consequently the widest horseshoe flow will over-shoot the horseshoe region, and exit after just one horseshoe turn. We call this the ``transient'' horseshoe flow.
\item[(4)] As the transient flow pushes into the disk, it is deflected by midplane material, resulting in a vortex roll-up (\figref{fig:rz}). At the same time, the loss of material in the horseshoe region due to the transient flow is replenished by the high altitude flow lying just outside of the region \figref{fig:mix}. This generates a meridional circulation that mixes the flow (\figref{fig:mix}).
\item[(5)] The meridional circulation (or the $\phi$-direction vortex) is eventually killed by disk viscosity, and the flow resets to the columnar flow in (1) before the next encounter with the planet.
\end{itemize}

The flow speed inside the Bondi sphere approaches $\cs$, so the gas density there is much less dense than if it has a hydrostatic structure (\figref{fig:atm}). Nearly all of the gas in the Bondi sphere participates in the horseshoe flow. We found that only gas within a distance of $\sim\rs$ is bound to the planet.

We also found that as a part of the asymmetry in the flow pattern across $r=a$, the stagnation points are now offset from the azimuth of the planet, $\phi_{\rm p}$, where the inner point now lies below $\phi_{\rm p}$ and the outer point above. The flow pattern asymmetry corresponds to an imbalance in the inner and outer horseshoe flow, which generates a net corotation torque of $\sim1.5T_0$. Overall, this results in a net torque of $\sim-0.77T_0$, much reduced from $-2.19T_0$ predicted by linear calculations.

\subsection{Forming Gaseous Planets}
\label{sec:grow}
Following OSK15 to measure the flux of mass in and out of the Bondi sphere by $t_{\rm replenish}$, we found $t_{\rm replenish}\sim \Omega_{\rm p}^{-1}$, which is alarmingly short if the Bondi sphere is the planet's atmosphere. Through streamline analysis, we found that nearly all streamlines within the Bondi sphere are not bound to the planet. It is then more appropriate to classify the Bondi sphere by its flow topology: a part of the transient horseshoe flow. However, this leaves us with little atmosphere. If this is universally true for all planetary cores, then gaseous planets with cores of a few $M_{\oplus}$, cannot be formed. 

There are two major issues with our result in the context of gas accretion. First, it should be reminded that we did find the gas within $1.5\rs$ of the planet to be bound, but it is only resolved by 3 grid cells. Increasing resolution in this region will allow us to be more certain about how much gas is truly bound to the planet. The resolution required to fully resolve the planetary atmosphere is of order the pressure scale height on the surface of the planet: $\cs^2 \rs^2 /(q G M_\ast) \sim 0.1\rs$ for our setup. This kind of resolution is attainable with local simulations around the planet.

Second, and more importantly, our simulation is globally isothermal, which is unrealistic since it does not take into account the heating and cooling of the atmosphere. A planet's atmosphere is expected to be heated through the accretion of gas and planetesimals, and the timescale for it to cool from that heat is much longer than $t_{\rm replenish}$ measured from our simulation; therefore the planet's atmosphere should be more appropriately described as adiabatic rather than isothermal. A more heated and pressurized atmosphere may deflect the transient horseshoe flow and resist it from intruding into the Bondi sphere, allowing more gas to be bound to the planet. This may have already been observed in the 3D radiation-hydrodynamics simulations by \citet{DAngelo2013}, which showed that the planet has a bound atmosphere of the size of its Bondi sphere; however, their relatively large disk viscosity (see \secref{sec:viscosity}) is expected to slow down vertical inflow speed and weaken 3D effects. It remains to be seen how large the bound atmosphere is in an inviscid flow with realistic radiative transfer.

\subsection{Stopping Type I Migration}
\label{sec:stop}
We have shown that our 3D disk produces a net corotation torque that is not expected in 2D analysis. As a result, our planet, embedded in a 3D disk, migrates 3 times slower than if it is driven by the differential Lindblad torque alone. This result is subject to a number of uncertainties. 

First, a major limitation to the accuracy of our torque measurement is the poor numerical accuracy near the planet (see \figref{fig:stag}). Because of this we have to exclude the $r=0.5\rB$ sphere around the planet from torque calculation. \citet{Ormel2015a} showed that numerical convergence can be more efficiently achieved if the grid geometry around the planet is polar rather than Cartesian. This is challenging to implement in a global simulation, because the grid should, ideally, also be polar around the star-planet's center of mass. Therefore we did not attempt it. Nevertheless, we believe our torque measurements do capture the essence of 3D effects, because our region of exclusion is sufficiently small that it does not cover the stagnation points, and we used an independent method to measure the corotation torque which gave a consistent result.

Second, we have chosen a disk profile that minimizes the net corotation torque in order to more easily identify differences between 2D and 3D. We have seen in \secref{sec:tor} that the one-sided corotation torque has a magnitude that can overwhelm the Lindblad torque, so an imbalance between the inner and outer horseshoe flow can potentially dominate type I migration rate. This can be accomplished by a modification as simple as changing the disk density or sound speed profile. A future study on how the net 3D corotation torque depends on disk parameters will be valuable to understanding planet migration.

Third, we have not considered thermal physics in our model, which has been shown to be capable of reversing type I migration \citep[e.g.][]{Bitsch2011,Bitsch2014,Lega2014,Masset2015}. Before considering the full radiative transfer problem, a possible first step will be to relax our isothermal condition to an adiabatic equation of state. 2D results \citep[][]{Masset2009,Paardekooper2010} have shown that an additional corotation torque related to the enthalpy of the fluid is expected for an adiabatic disk. The 3D aspect of this should be studied in more detail.

Fourth, our planet has a fixed circular orbit, despite the fact we are measuring a non-zero net torque on it. From our results , we can speculate that when the planet migrates inward (outward), disk material may be transported from the inner (outer) to the outer (inner) disk via the transient horseshoe flow, which may further unbalance the inner and outer flow and generate a larger net torque on the planet, leading to type III migration \citep{typeIII}. The 3D dynamical interaction between the planet and the disk has the potential to modify calculations based on fixed planets by a margin as large as the one-sided corotation torque.

Finally, we note that our planet has a relatively large mass comparing to planets that typically fall in the type I regime, which are $\lesssim1~M_{\oplus}$. This raises the question of how well our torque measurement applies to type I migration in general. We believe the horseshoe flow asymmetry that causes the stagnation point offset is applicable to lower mass planets, because the flow pattern we presented shares many similarities with OSK15's (compare our \figref{fig:mid} to their Figure 3), who simulated the inviscid flow around a planet with $q_{\rm th}=0.01$. Therefore, lower mass planets should also experience a reduced migration rate like ours. However, it is unclear whether the magnitude of the reduction will be similar. This question will be answered when global inviscid 3D simulations of smaller planets become feasible. 

The ability to simulate smaller planets in 3D is very dependent on computational resources. If we attempt to simulate a $1~M_{\oplus}$ planet, then in order to have the same resolution across $\rB$ as we do in this paper, cell sizes would need to decrease by a factor of 5, and the computational time scales as $5^{3+1}=625$, where the power of 3 comes from the increase in the total number of cells, and 1 from the shortened timestep due to the Courant limit. This amount of computational resources is much beyond what is currently available to us, but in the future, advancements in both hardware, such as access to a large GPU cluster, and software, such as a more sophisticated grid design, may open this pathway for us.

\subsection{Torque and Viscosity}
\label{sec:saturation}
In \secref{sec:viscosity} we suggested to use the condition $t_{\nu}>t_{\rm lib}$ to identify where 3D effects become important to the flow topology and planetary torque. This is also the condition for torque saturation in 2D \citep{Ward1991}. Since we do measure a non-zero net corotation torque, it is important to ask whether this torque will saturate. Evidently, \figref{fig:time} does not show any sign of torque saturation.

The process of torque saturation can be described as follows: because horseshoe orbits are closed in 2D, the entire horseshoe region is completely separated from the rest of the disk, and therefore has only a finite amount of angular momentum to exchange with the planet. Consequently, the net exchange of angular momentum between the horseshoe region and the planet in steady state must be zero. The corotation torque can only be unsaturated if fresh supply of angular momentum enters the horseshoe region, which can be done through viscous diffusion, hence the torque saturation condition. In \secref{sec:tur_flow} we showed that there is a constant exchange of material between the horseshoe region and the rest of the disk due to the existence of the transient horseshoe flow. The planet is therefore able to replenish the horseshoe flow without help from disk viscosity. In other words, when $t_{\nu}>t_{\rm lib}$, 3D effects kick in, and the corotation torque is unsaturated; when $t_{\nu}<t_{\rm lib}$, disk viscosity dominates, and it is also unsaturated. So corotation torque saturation may be a pure 2D effect. For a future study, it would be interesting to test a range of disk profiles and viscosity in 3D, and find out under what circumstance does torque saturation occur.

\section*{Acknowledgments}

We thank Eugene Chiang, Fr\'{e}d\'{e}ric Masset, and Chris Ormel for helpful feedback that substantially improved this manuscript. We also thank Ruth Murray-Clay for insightful discussions during the early stages of this work. We gratefully acknowledge support from the Discovery Grant by the Natural Sciences and Engineering Research Council of Canada.

\appendix
\section{Radial Flow Speed in the Transient Horseshoe Flow}
\label{append}

In this appendix we analytically calculate the radial outflow speed at which the transient horseshoe flow exits the horseshoe region. Bernoulli's constant for a given streamline can be written as:
\begin{equation}
B = -\frac{1}{2} r^2 \Omega_{\rm p}^2 + \Phi + \frac{1}{2} |\bm{u}|^2 + \eta \, .
\label{eqn:bern}
\end{equation}
We divide $\eta$ into two components: $\eta = \eta_0+\eta_{\rm p}$, where $\eta_0$ is the background enthalpy profile that balances the star's potential:
\begin{equation}
\frac{{\rm d}\eta_0}{{\rm d}z} = -\frac{\rm d}{{\rm d}z} \frac{GM_\ast}{\sqrt{r^2+z^2}} \, .
\label{eqn:eta0}
\end{equation}
For $x = r-a$ where $x\ll a$, we can rewrite \eqnref{eqn:bern} as:
\begin{equation}
B = -\frac{3}{2} a^2 \Omega_{\rm p}^2 - \frac{3}{2} x^2 \Omega_{\rm p}^2 - q\frac{G M_\ast}{\sqrt{x^2+y^2+z^2}} + \frac{1}{2} |\bm{u}|^2 + \eta_{\rm p} \, ,
\label{eqn:bern2}
\end{equation}
where $\{x,y,z\}$ are the local Cartesian coordinates centering on the planet. In \eqnref{eqn:bern2}, $\eta_0$ has been canceled by the z-dependent part of the star's potential. We will also drop $\eta_{\rm p}$ because it is expected to have a small contribution to \eqnref{eqn:bern2} comparing to the planet's gravitational potential, because as shown in \figref{fig:atm}, the density, hence pressure, near the planet is much less than what is needed to balance the planet's gravity.

Consider two points along an inner horseshoe orbit: the first point is just before the fluid is about to perform its horseshoe turn, located at $\bm{p}_{1}=\{x_1,y_1,z_1\}$, where $x_1<0$ , and its velocity can be approximated as $\bm{u}=\{u_{\rm r,1,2},\frac{3}{2}x_1 \Omega_{\rm p},0\}$, where $\frac{3}{2}x_1 \Omega_{\rm p}$ is the local Keplerian shear; the second point is after the turn, having been pulled down to the midplane, at $\bm{p}_{2}=\{x_2,y_2,0\}$, where $x_2=-x_1$ and $y_1=y_2$, and has $\bm{u}=\{u_{\rm r,2},\frac{3}{2}x_2 \Omega_{\rm p},0\}$. For convenience we define $d^2 = x_1^2+y_1^2 = x_2^2+y_2^2$ as the distance from the planet on the $x-y$ plane. Equating $B$ at these two points, we have:
\begin{equation}
-q\frac{G M_\ast}{\sqrt{d^2+z^2}} + \frac{1}{2}u_{\rm r,1}^2 = -q\frac{G M_\ast}{d} + \frac{1}{2}u_{\rm r,2}^2 \, .
\label{eqn:ur1}
\end{equation}
By our results in \secref{sec:cor_flow}, it is safe to assume all of the widest horseshoe orbits has the same $x_1$ and $y_1$ irrespective of their starting $z$. This allows us to approximate $z^2\approx h_0^2$ in \eqnref{eqn:ur1} since:
\begin{equation}
\frac{\int_0^\infty z^2 \rho(z=0)e^{-\frac{z^2}{2 h^2}} {\rm d}z}{\int_0^\infty \rho(z=0)e^{-\frac{z^2}{2 h^2}} {\rm d}z} = h^2 \, ,
\label{eqn:z2}
\end{equation}
Finally, combining Equations \ref{eqn:ur1} and \ref{eqn:z2}, we have:
\begin{equation}
|u_{\rm r,2}| = \sqrt{2q\frac{GM}{d}\left(1-\frac{d}{\sqrt{d^2+h_0^2}}\right)+u_{\rm r,1}^2} \, .
\label{eqn:ur2}
\end{equation}
Since $|u_{\rm r,2}| > |u_{\rm r,1}|$, the horseshoe streamline is asymmetric about $x=0$. \eqnref{eqn:ur2} should be considered an upper limit for two main reasons: first, the streamlines in fact only fall to a fraction of $h_0$ instead the midplane; and second, an increase in enthalpy as a fluid element moves toward the midplane can compensate for some loss in gravitational potential energy. In fact if $\eta$ satisfies \eqnref{eqn:balance}, it would leave $|u_{\rm r,1}|=|u_{\rm r,2}|$, resulting in a symmetric horseshoe flow. \eqnref{eqn:ur2} is useful however, for allowing us to estimate how this outflow speed scales with planet mass. 

$d$ should scale with the horseshoe half-width $w$, which can have two possible scalings: $w\propto a\sqrt{q(a/h_0)}$ for low mass planets \citep{Masset2006, Paardekooper2009}, and $w\propto \rH$ for high mass planets \citep{Masset2006, Peplinski2008}. In the low mass limit, if we approximate $d\sim a\sqrt{q(a/h_0)}$, and further simplify \eqnref{eqn:ur2} using $d\ll h$ (equivalently, $q_{\rm th}\ll1$), then we get:
\begin{equation}
|u_{\rm r,2}| \simeq \sqrt{ 2\sqrt{q_{\rm th}} \cs^2 +  |u_{\rm r,1}|^2} \, .
\label{eqn:ur_low}
\end{equation}
In the high mass limit, we approximate $d\sim 2\rH$, and have $\rH\gg h$ (equivalently, $q_{\rm th}\gg1$):
\begin{equation}
|u_{\rm r,2}| \simeq \sqrt{ \frac{3}{8}\cs^2 +  |u_{\rm r,1}|^2} \, .
\label{eqn:ur_high}
\end{equation}
These two limits show that the injection of energy due to the change in $z$ is smaller if planet mass decreases, and asymptotes to a constant value for a high mass planet. In our simulation, the radial velocity measured at $|x|=\rH$ ranges from $0.2$ to $0.4\cs$, depending on the azimuth at which it is measured (see Figures \ref{fig:surf} and \ref{fig:rz}). For comparison, \eqnref{eqn:ur_high} predicts a velocity of $\sim 0.6~\cs$ if $|u_{\rm r,1}|\ll\cs$, which is within an order of unity from our numerical result.

\bibliographystyle{apj}
\bibliography{Lit}

\begin{thebibliography}{}
\expandafter\ifx\csname natexlab\endcsname\relax\def\natexlab#1{#1}\fi

\bibitem[{{Artymowicz}(1993)}]{Artymowicz1993}
{Artymowicz}, P. 1993, \apj, 419, 166

\bibitem[{{Ayliffe} \& {Bate}(2012)}]{Ayliffe2012}
{Ayliffe}, B.~A., \& {Bate}, M.~R. 2012, \mnras, 427, 2597

\bibitem[{{Batalha} {et~al.}(2013){Batalha}, {Rowe}, {Bryson}, {Barclay},
  {Burke}, {Caldwell}, {Christiansen}, {Mullally}, {Thompson}, {Brown},
  {Dupree}, {Fabrycky}, {Ford}, {Fortney}, {Gilliland}, {Isaacson}, {Latham},
  {Marcy}, {Quinn}, {Ragozzine}, {Shporer}, {Borucki}, {Ciardi}, {Gautier},
  {Haas}, {Jenkins}, {Koch}, {Lissauer}, {Rapin}, {Basri}, {Boss}, {Buchhave},
  {Carter}, {Charbonneau}, {Christensen-Dalsgaard}, {Clarke}, {Cochran},
  {Demory}, {Desert}, {Devore}, {Doyle}, {Esquerdo}, {Everett}, {Fressin},
  {Geary}, {Girouard}, {Gould}, {Hall}, {Holman}, {Howard}, {Howell},
  {Ibrahim}, {Kinemuchi}, {Kjeldsen}, {Klaus}, {Li}, {Lucas}, {Meibom},
  {Morris}, {Pr{\v s}a}, {Quintana}, {Sanderfer}, {Sasselov}, {Seader},
  {Smith}, {Steffen}, {Still}, {Stumpe}, {Tarter}, {Tenenbaum}, {Torres},
  {Twicken}, {Uddin}, {Van Cleve}, {Walkowicz}, \& {Welsh}}]{Batalha2013}
{Batalha}, N.~M., {Rowe}, J.~F., {Bryson}, S.~T., {et~al.} 2013, \apjs, 204, 24

\bibitem[{{Ben{\'{\i}}tez-Llambay} {et~al.}(2015){Ben{\'{\i}}tez-Llambay},
  {Masset}, {Koenigsberger}, \& {Szul{\'a}gyi}}]{Masset2015}
{Ben{\'{\i}}tez-Llambay}, P., {Masset}, F., {Koenigsberger}, G., \&
  {Szul{\'a}gyi}, J. 2015, \nat, 520, 63

\bibitem[{{Bitsch} \& {Kley}(2011)}]{Bitsch2011}
{Bitsch}, B., \& {Kley}, W. 2011, \aap, 530, A41

\bibitem[{{Bitsch} {et~al.}(2014){Bitsch}, {Morbidelli}, {Lega}, \&
  {Crida}}]{Bitsch2014}
{Bitsch}, B., {Morbidelli}, A., {Lega}, E., \& {Crida}, A. 2014, \aap, 564,
  A135

\bibitem[{{Blondin} \& {Lufkin}(1993)}]{VH1}
{Blondin}, J.~M., \& {Lufkin}, E.~A. 1993, \apjs, 88, 589

\bibitem[{{Borucki} {et~al.}(2010){Borucki}, {Koch}, {Basri}, {Batalha},
  {Brown}, {Caldwell}, {Caldwell}, {Christensen-Dalsgaard}, {Cochran},
  {DeVore}, {Dunham}, {Dupree}, {Gautier}, {Geary}, {Gilliland}, {Gould},
  {Howell}, {Jenkins}, {Kondo}, {Latham}, {Marcy}, {Meibom}, {Kjeldsen},
  {Lissauer}, {Monet}, {Morrison}, {Sasselov}, {Tarter}, {Boss}, {Brownlee},
  {Owen}, {Buzasi}, {Charbonneau}, {Doyle}, {Fortney}, {Ford}, {Holman},
  {Seager}, {Steffen}, {Welsh}, {Rowe}, {Anderson}, {Buchhave}, {Ciardi},
  {Walkowicz}, {Sherry}, {Horch}, {Isaacson}, {Everett}, {Fischer}, {Torres},
  {Johnson}, {Endl}, {MacQueen}, {Bryson}, {Dotson}, {Haas}, {Kolodziejczak},
  {Van Cleve}, {Chandrasekaran}, {Twicken}, {Quintana}, {Clarke}, {Allen},
  {Li}, {Wu}, {Tenenbaum}, {Verner}, {Bruhweiler}, {Barnes}, \&
  {Prsa}}]{Kepler}
{Borucki}, W.~J., {Koch}, D., {Basri}, G., {et~al.} 2010, Science, 327, 977

\bibitem[{{Colella} \& {Woodward}(1984)}]{PPM}
{Colella}, P., \& {Woodward}, P.~R. 1984, Journal of Computational Physics, 54,
  174

\bibitem[{{D'Angelo} \& {Bodenheimer}(2013)}]{DAngelo2013}
{D'Angelo}, G., \& {Bodenheimer}, P. 2013, \apj, 778, 77

\bibitem[{{D'Angelo} {et~al.}(2003){D'Angelo}, {Kley}, \&
  {Henning}}]{DAngelo2003}
{D'Angelo}, G., {Kley}, W., \& {Henning}, T. 2003, \apj, 586, 540

\bibitem[{{D'Angelo} \& {Lubow}(2010)}]{DAngelo2010}
{D'Angelo}, G., \& {Lubow}, S.~H. 2010, \apj, 724, 730

\bibitem[{{Fung} {et~al.}(2014){Fung}, {Shi}, \& {Chiang}}]{Fung2014}
{Fung}, J., {Shi}, J.-M., \& {Chiang}, E. 2014, \apj, 782, 88

\bibitem[{{Ikoma} {et~al.}(2000){Ikoma}, {Nakazawa}, \& {Emori}}]{Ikoma2000}
{Ikoma}, M., {Nakazawa}, K., \& {Emori}, H. 2000, \apj, 537, 1013

\bibitem[{{Kley}(1998)}]{Kley98}
{Kley}, W. 1998, \aap, 338, L37

\bibitem[{{Lee} {et~al.}(2014){Lee}, {Chiang}, \& {Ormel}}]{Lee2014}
{Lee}, E.~J., {Chiang}, E., \& {Ormel}, C.~W. 2014, \apj, 797, 95

\bibitem[{{Lega} {et~al.}(2014){Lega}, {Crida}, {Bitsch}, \&
  {Morbidelli}}]{Lega2014}
{Lega}, E., {Crida}, A., {Bitsch}, B., \& {Morbidelli}, A. 2014, \mnras, 440,
  683

\bibitem[{{Masset} \& {Casoli}(2009)}]{Masset2009}
{Masset}, F.~S., \& {Casoli}, J. 2009, \apj, 703, 857

\bibitem[{{Masset} \& {Casoli}(2010)}]{Masset2010}
---. 2010, \apj, 723, 1393

\bibitem[{{Masset} {et~al.}(2006){Masset}, {D'Angelo}, \& {Kley}}]{Masset2006}
{Masset}, F.~S., {D'Angelo}, G., \& {Kley}, W. 2006, \apj, 652, 730

\bibitem[{{Masset} \& {Ogilvie}(2004)}]{Masset2004}
{Masset}, F.~S., \& {Ogilvie}, G.~I. 2004, \apj, 615, 1000

\bibitem[{{Morbidelli} {et~al.}(2014){Morbidelli}, {Szul{\'a}gyi}, {Crida},
  {Lega}, {Bitsch}, {Tanigawa}, \& {Kanagawa}}]{Morbidelli2014}
{Morbidelli}, A., {Szul{\'a}gyi}, J., {Crida}, A., {et~al.} 2014, \icarus, 232,
  266

\bibitem[{{Ormel}(2013)}]{Ormel2013}
{Ormel}, C.~W. 2013, \mnras, 428, 3526

\bibitem[{{Ormel} {et~al.}(2015{\natexlab{a}}){Ormel}, {Kuiper}, \&
  {Shi}}]{Ormel2015a}
{Ormel}, C.~W., {Kuiper}, R., \& {Shi}, J.-M. 2015{\natexlab{a}}, \mnras, 446,
  1026

\bibitem[{{Ormel} {et~al.}(2015{\natexlab{b}}){Ormel}, {Shi}, \&
  {Kuiper}}]{Ormel2015b}
{Ormel}, C.~W., {Shi}, J.-M., \& {Kuiper}, R. 2015{\natexlab{b}}, \mnras, 447,
  3512

\bibitem[{{Paardekooper} {et~al.}(2010){Paardekooper}, {Baruteau}, {Crida}, \&
  {Kley}}]{Paardekooper2010}
{Paardekooper}, S.-J., {Baruteau}, C., {Crida}, A., \& {Kley}, W. 2010, \mnras,
  401, 1950

\bibitem[{{Paardekooper} {et~al.}(2011){Paardekooper}, {Baruteau}, \&
  {Kley}}]{Paardekooper2011}
{Paardekooper}, S.-J., {Baruteau}, C., \& {Kley}, W. 2011, \mnras, 410, 293

\bibitem[{{Paardekooper} \& {Papaloizou}(2008)}]{PaardekooperP2008}
{Paardekooper}, S.-J., \& {Papaloizou}, J.~C.~B. 2008, \aap, 485, 877

\bibitem[{{Paardekooper} \& {Papaloizou}(2009)}]{Paardekooper2009}
---. 2009, \mnras, 394, 2297

\bibitem[{{Papaloizou} {et~al.}(2007){Papaloizou}, {Nelson}, {Kley}, {Masset},
  \& {Artymowicz}}]{typeIII}
{Papaloizou}, J.~C.~B., {Nelson}, R.~P., {Kley}, W., {Masset}, F.~S., \&
  {Artymowicz}, P. 2007, Protostars and Planets V, 655

\bibitem[{{Peplinski}(2008)}]{Peplinski2008}
{Peplinski}, A. 2008, PhD thesis, University of Stockholm

\bibitem[{{Petigura} {et~al.}(2013){Petigura}, {Marcy}, \&
  {Howard}}]{Petigura2013}
{Petigura}, E.~A., {Marcy}, G.~W., \& {Howard}, A.~W. 2013, \apj, 770, 69

\bibitem[{{Pollack} {et~al.}(1996){Pollack}, {Hubickyj}, {Bodenheimer},
  {Lissauer}, {Podolak}, \& {Greenzweig}}]{Pollack1996}
{Pollack}, J.~B., {Hubickyj}, O., {Bodenheimer}, P., {et~al.} 1996, \icarus,
  124, 62

\bibitem[{{Rafikov}(2006)}]{Rafikov2006}
{Rafikov}, R.~R. 2006, \apj, 648, 666

\bibitem[{{Shakura} \& {Sunyaev}(1973)}]{alpha}
{Shakura}, N.~I., \& {Sunyaev}, R.~A. 1973, \aap, 24, 337

\bibitem[{{Tanaka} {et~al.}(2002){Tanaka}, {Takeuchi}, \& {Ward}}]{Tanaka2002}
{Tanaka}, H., {Takeuchi}, T., \& {Ward}, W.~R. 2002, \apj, 565, 1257

\bibitem[{{Tanigawa} {et~al.}(2012){Tanigawa}, {Ohtsuki}, \&
  {Machida}}]{Tanigawa2012}
{Tanigawa}, T., {Ohtsuki}, K., \& {Machida}, M.~N. 2012, \apj, 747, 47

\bibitem[{{Ward}(1991)}]{Ward1991}
{Ward}, W.~R. 1991, in Lunar and Planetary Science Conference, Vol.~22, Lunar
  and Planetary Science Conference, 1463

\end{thebibliography}
\end{document}